\documentclass[letterpaper,12pt]{article}
\usepackage{tabularx} 
\usepackage{amsmath,amssymb}  
\usepackage{graphicx} 
\usepackage[margin=1in]{geometry} 
\usepackage{cite} 
\usepackage[final]{hyperref} 
\usepackage[figurename=Figure, font=small]{caption}

\hypersetup{
	colorlinks=true,       
	linkcolor=blue,        
	citecolor=blue,        
	filecolor=magenta,     
	urlcolor=blue         
}

\def\trace{\operatorname{Tr}}
\def\gyr{\mathrm{gyr}}
\def\ni{0}
\def\kh{\mathrm{Kf}}
\def\bfa{\mathbf{a}}
\def\bfb{\mathbf{b}}

\def\bfd{\mathbf{d}}
\def\bfe{\mathbf{e}}

\def\bfh{\mathbf{h}}
\def\bfv{\mathbf{v}}

\def\bfD{\mathbf{D}}

\def\bbN{\mathbb{N}}
\def\bbZ{\mathbb{Z}}

\def\bbR{\mathbb{R}}

\def\cD{\mathcal{D}}

\def\cF{\mathcal{F}}
\def\cI{\mathcal{I}}
\def\cM{\mathcal{M}}
\def\cN{\mathcal{N}}
\def\cO{\mathcal{O}}
\def\cP{\mathcal{P}}
\def\cS{\mathcal{S}}
\def\cV{\mathcal{V}}

\title{\Large From Matrix Models to Gaussian Molecules and the Einstein-Hilbert Action}
\author{\normalsize Manfred Herbst\footnote{\small \href{mailto:maherbst@gmx.de}{maherbst@gmx.de}}\\
\textit{\small Institut für Theoretische Physik, Technische Universität Wien,}\\
\textit{\small Wiedner Hauptstraße 8-10/136, A-1040 Vienna, Austria}\\
}
\date{} 

\begin{document}

\maketitle

\begin{abstract}
A matrix model on a $D$-dimensional Euclidean space is introduced as a generalization of random matrix models and as a non-perturbative definition of discretized closed string theory. The free energy of the matrix model is formally derived to all orders in string perturbation expansion and shown to be given in terms of invariant graph polynomials, whose coefficients enumerate ribbon graphs and are a refinement of the generalized Catalan numbers. The vacuum diagrams contributing to the free energy are found to be related to Gaussian molecules, known from the study of polymer structures.

Coupling the matrix field to a curved background with Riemannian metric yields a non-perturbative definition of discretized string theory on this background. No on-shell condition for the metric is required to arrive at the free energy. Rather, it is shown that the free energy of the matrix model is the Einstein-Hilbert action with cosmological constant term. The gravitational and the cosmological constants are both formally determined to all orders in the string perturbation expansion. In fact, they are explicitly given by the expectation value of a particular graph invariant.

Introducing a vector field, minimally coupled to a background gauge field, provides a discretized open-closed string theory, leading to the Yang-Mills action as well as intrinsic and extrinsic curvature terms.
\end{abstract}

\section{Introduction and overview}

The subject of random matrix models is well-studied in the literature with a plethora of relations and applications to mathematics and physics, too many to enumerate and acknowledge. The story relevant for this work started with 't Hooft's groundbreaking work \cite{tHooft:1973alw} that intended to relate quantum chromodynamics for color gauge group $U(N)$ in the limit of many colors, $N \rightarrow \infty$, to a dual string theory. The two-dimensional surfaces of the latter are generated in a discrete manner through ribbon graph diagrams.

On the mathematical side the census of planar ribbon graph diagrams started earlier in a series of works by Tutte \cite{Tutte:1962-1968}. The first full census of ribbon graphs of arbitrary genus goes back to Walsh and Lehman \cite{Walsh:1972-1975}. They used combinatorial techniques to derive a recursion relation for enumerating ribbon graphs.

Brézin, Itzykson, Parisi and Zuber picked up 't Hooft's idea \cite{Brezin:1977sv} and started to exploit matrix model integrals for ribbon graph counting. They succeeded in enumerating all planar graphs by computing the generating function from a matrix model using a saddle point method. In subsequent work various methods were developed to evaluate higher genus contributions to the generating function, like orthogonal polynomials \cite{Bessis:1980ss}, loop equations and Virasoro constraints \cite{David:1990ge,GinspargZinnJustin:1990,GrossMigdal:1990,Kazakov:1989bc,Mironov:1990im,David:1990sk,Dijkgraaf:1990rs,Fukuma:1990jw,Makeenko:1990in} as well as topological recursion \cite{EynardOrantin:2007}.
In fact, for the Hermitian one-matrix model, it was shown in \cite{Mulase:2012yk,Dumitrescu:2012dka} that topological recursion is equivalent to the recursive equations for ribbon graph counting developed by Walsh and Lehman \cite{Walsh:1972-1975}.

The relation of matrix models to 2d gravity and non-critical string theory, that is Liouville field theory coupled to matter fields, is well-studied for $D \leq 1$ \cite{Kazakov:1988ch,Brezin:1990rb}. Collective field theory techniques showed that the Liouville mode is generated from the matrix degrees of freedom in the continuum limit \cite{Das:1990kaa,Kostov:1999xi}, see also \cite{Ginsparg:1993is} and references therein.
 
Although there have been hints and remarks on how to go beyond $D=1$ for matrix models, see for instance \cite{Kazakov:1988ch,Boulatov:1986jd}, this route was not taken, one obstacle being that the relation to non-critical string theory and Liouville theory in the double scaling limit is unclear in the window $1<D<25$, since according to Liouville theory the critical exponent is then complex-valued \cite{Knizhnik:1988ak, David:1988hj, Distler:1988jt, Ginsparg:1993is}.
Another reason is that the techniques to solve matrix models for $D \leq 1$ cease to work for $D > 1$.

This work will not give any insights into this relation either, but will derive the free energy for the Hermitian one-matrix model as a formal power series in arbitrary dimension $D\geq 0$ and in curved background. The motivation for doing this is two-fold. On the one hand, we expect to find new graph combinatorics in terms of invariant graph polynomials related to the generalized Catalan numbers. On the other hand, we couple the matrix model to a non-trivial, \emph{a priori} unconstrained, metric background and study the effect on the metric dynamics itself.

In formulating the matrix model in higher dimensions some properties shall be assumed:

First, the matrix model is a Hermitian matrix model with a single $N\times N$ matrix field $M=M(x)$ on a manifold $\cM$, where initial focus will be on $\cM=\bbR^D$. Whenever a metric is introduced to measure distances, it shall be of Euclidean signature, that is $\cM$ is a Riemannian manifold. Further restrictions on $\cM$ will be given when needed. 
	
Second, the matrix $M$ is subject to the potential
$$
  V(M) = \sum_{d=1}^\infty 
	t_d \trace{M^d}\, .
$$
The free energy of the matrix model will be treated as a formal power series in the parameters $t_d$. For some considerations we will restrict attention to a monomial potential, foremost $V(M) = t_4 \trace{M^4}$.

And third, the functional integral representation of the higher-dimensional matrix model shall not show infinities or singularities other than the ones related to the infinite volume of $\cM$ and to the convergence of the perturbation expansion. No regularization shall be required (just as for the integral representation of random matrix models for $D < 1$).

A naive attempt to formulate the matrix model in $D \geq 1$ solely based on the potential $V(M)$, does lead to an ill-defined functional integral. This can be seen when trying to perturbatively solve the integral around the quadratic potential term, $\int d^Dx \trace{M^2}$. The latter formally corresponds to a propagator given by a Dirac delta distribution $\delta^D(x-x')$. Working out Feynman diagrams for a graph of $v$ vertices and $e$ edges would lead to integrals of the form
$$
  \sim \int d^Dx_1 \ldots \int d^Dx_v \prod_{e_{ij}=1}^e \delta^D(x_i-x_j)\, ,
$$
which are ill-defined, because generally $e > v$. 

The third point therefore requires introducing a kinetic term for the matrix model that gives rise to a propagator and to Feynman diagrams that do not require regularization. This excludes the local quadratic kinetic term in terms of the Laplacian operator $\Delta$.
Instead, by introducing a small parameter $\alpha$ that sets a length scale $\sqrt{\alpha}$, the propagator for the kinetic term shall reproduce the Dirac delta distribution for vanishing $\alpha$. The simplest and natural choice for the propagator is the Gaussian distribution. Its role as heat kernel for the heat equation suggests considering the non-local kinetic term 
$e^{-\frac{\alpha}{2}\Delta}$. In fact, the heat kernel is at the same time the Greens function for this kinetic term and the solution to the heat equation.

In this work we study the higher-dimensional matrix model
\begin{eqnarray}
	\label{eq:higherdim_matrix_model}
	Z(\alpha, \lambda, t_d) &=& e^{-\cF(\alpha,\lambda,t_d)} =
	\int{\cD M(x)\; e^{-S(\alpha, \lambda, t_d, M(x))}}\, , \\
	S(\alpha, \lambda, t_d, M(x)) &=& 
	\frac{N}{(2\pi\alpha)^{D/2}} \int_\cM {d^Dx \; 
	\left(\frac{1}{2\lambda} 
	\trace M(x) e^{-\frac{\alpha}{2}\Delta_x} M(x) + 
	V(M)\right)}\, , \nonumber
\end{eqnarray}
as a non-perturbative, discrete definition of a string theory.
The coupling $\lambda$ controls the weight between the kinetic and potential term. The parameter $\lambda$ and the parameters $t_d$ are redundant, one of them could be removed by a field redefinition. 

The first main result is related to graph combinatorics. 
We are interested in deriving the free energy $\hat\cF$ of the matrix model as a formal power series in $\lambda$ and $t_d$. For $\cM = \bbR^D$ and Euclidean metric, we find that the free energy takes the form
\begin{equation}
	\label{eq:free_energy_flat}
  \hat\cF = \frac{V_\cM}{(2\pi\alpha)^{D/2}}\, \hat w(\lambda,t_d,N)\ ,
\end{equation}
where $\alpha$ and the (infinite) volume $V_\cM$ only appear in the  prefactor. The formal power series $\hat w$ encodes all the graph combinatorics and enumeration in terms of invariant graph polynomials $\cP_{\bfd,\gamma}(N^{-2})$ for skeleton graphs $\gamma$ with vertex degrees $\bfd = (d_1, \ldots, d_v)$.
The coefficients of the graph polynomials turn out to be a refinement of the generalized Catalan numbers of \cite{Walsh:1972-1975,Mulase:2012yk}, enumerating the number of ribbon graphs associated with the skeleton $\gamma$ (cf. Figure \ref{fig:RibbonGraphsSkeleton}).

For consistency reasons (particularly when introducing a background metric in the following) it is important to study the size of vacuum bubbles of the matrix model, that is the size of graph embeddings into $\cM$. Vacuum bubbles of the matrix model turn out to be described by Gaussian molecules, a formulation of molecules through Gaussian interactions, which is used to study structural and physical properties of molecules such as polymers. In particular, a well-known result from the theory of Gaussian molecules going back to \cite{Eichinger:1980} relates the gyration radius of a molecule with a graph invariant that allows an explicit estimation of its size. Applied to the matrix model we get a handle on estimating the growth of the size of vacuum bubbles with the vertex number, cf. \cite{Boulatov:1986jd}.

For the second main result, the matrix integral is considered on a Riemannian manifold $(\cM,g)$ with metric $g$, which is assumed to be simply connected and $\sqrt{\alpha}$ being much smaller than the size of 
$\cM$. The functional integral (\ref{eq:higherdim_matrix_model}) must be written in covariant form, in particular, the Laplacian operator is the covariant Laplacian with respect to the Riemannian metric $g$. No on-shell condition is required to derive the leading $\alpha$ approximation of the free energy. It turns out to be the Einstein-Hilbert action with cosmological constant,
\begin{equation}
	\label{eq:free_energy_curved}
  \hat \cF = \frac{1}{(2\pi\alpha)^{D/2}}
	\int d^Dx \sqrt{g}\,
	\hat w \left(
	  1 + \frac{\alpha}{12} 
		\langle\, \hat\cV\, \rangle_{\hat w}\,  R
	\right) .
\end{equation}
Here, $\hat w$ is the formal power series from (\ref{eq:free_energy_flat}), and the coefficient $\alpha\, \langle\, \hat\cV\, \rangle_{\hat w}$ of the curvature term is the expectation value of a graph invariant, explicitly given in (\ref{eq:total_graph_invariant}). In fact, in the continuum limit it is a measure for the average number of interactions in a vacuum bubble and is related to the cosmological constant by
$$
  \Lambda^{-1} = \frac{\alpha}{6}\, \langle\, \hat\cV\, \rangle_{\hat w} \, .
$$

We close this introduction with a short outline of the structure of this article. It starts with a review of the Hermitian one-matrix model in Section \ref{sec:HOMM}, focusing on graph enumeration encoded in the free energy and the Walsh-Lehman recursion relations.

In Section \ref{sec:MMAD} the Hermitian one-matrix model (\ref{eq:higherdim_matrix_model}) is considered for $D > 0$. The heat kernel for the Laplacian operator $\Delta_x$ is used to write the matrix model free energy $\hat\cF$ as perturbative expansion of the discretized string world sheet. Furthermore, the discrete string embedding coordinates can be integrated out, and the free energy $\hat\cF$ is derived as generating function for invariant graph polynomials. 
In Section \ref{sec:SchwingerDyson} the simplest Schwinger-Dyson equation is verified to be satisfied by the free energy (\ref{eq:free_energy_flat}). Section \ref{sec:GraphComplexity} discusses the appearance of the spanning tree entropy in the free energy, a well-studied property in graph combinatorics that measures the graph complexity. Section \ref{sec:Gyration} is dedicated to investigating the size of a vacuum bubble in the matrix model through the gyration radius of Gaussian molecules. The continuum limit is discussed in Section \ref{sec:ContinuumLimit}.

In Section \ref{sec:MMCB} the Hermitian one-matrix model is studied in a curved background with Riemannian metric. It is shown that the leading contribution to the free energy is the Einstein-Hilbert action (\ref{eq:free_energy_curved}) with cosmological constant.

In Section \ref{sec:VMM} a vector matrix model is considered to make contact with a discrete version of open-closed string theory. As shown in Section \ref{sec:MMGF}, introducing a minimally coupled gauge connection in the Laplacian operator for the vector matrix model leads to a free energy with effective non-Abelian Yang-Mills action on a membrane world volume. Section \ref{sec:covariant_generalization} is dedicated to a covariant generalization, including both a curved and gauge field background, which is shown to result in extrinsic and intrinsic curvature terms on the membane world volume.

Section \ref{sec:concluding_remarks} provides concluding remarks and outlines several aspects and questions that are not addressed in this article.\\[3pt]
\noindent {\bf Acknowledgements:} The author thanks Anton Rebhan, Harald Skarke, Wolfgang Lerche, Shota Komatsu, Johanna Knapp, and Johannes Walcher for their encouragement, support, and constructive comments on the manuscript. 

\section{The Hermitian one-matrix model}
\label{sec:HOMM}

In this section the stage will be set with selected properties of the Hermitian one-matrix model. The matrix integral is defined by (\ref{eq:higherdim_matrix_model}) for $D=0$. The integration measure is chosen with the normalization as 
$$
  \cD M = (2\pi)^{-N^2/2} \cdot 
	\prod_{i<j} d M^{\mathrm{Re}}_{ij} 
	\prod_{i<j} d M^{\mathrm{Im}}_{ij} 
	\prod_i M_{ii}\, ,
$$
where $M_{ij} = \frac{1}{\sqrt{2}}(M^{\mathrm{Re}}_{ij} + i M^{\mathrm{Im}}_{ij})$ for $i<j$.

To compute the partition function perturbatively using Feynman diagrams, the matrix $M$ is coupled to a source matrix $J$,
$$
	Z(J) = \int{\cD M\; e^{-S + \trace\left(J M\right)}}.
$$
Using quadratic completion and field redefinition this can be rewritten as
\begin{equation*}
	Z = Z_{\ni } \left. 
	\exp\left(-N V \left(\frac{\delta}{\delta J}\right)\right)
	\exp\left(\frac{\lambda}{2N} \trace J^2\right)\right|_{J=0},
\end{equation*}
where $Z_{\ni }$ is the Gaussian part of the partition function,
$$
  Z_{\ni } = \int{\cD M\; \exp{\left(-\frac{N}{2\lambda}	\trace M^2 \right)}} = \left(\frac{\lambda}{N}\right)^{\frac{N^2}{2}}
	\, .
$$
Its contribution to the free energy is therefore
$$
  \cF_{\ni } = - \ln Z_{\ni } = \frac{N^2}{2} \ln\left(N/\lambda\right) \, .
$$

For the interacting part we consider the normalized partition function $\hat Z = Z/Z_{\ni }$ and the free energy $\hat \cF = - \ln{\hat Z}$. Applying Feynman rules, the free energy is the sum over all connected Feynman diagrams. In particular, all contributions with $v$ vertices come from
$$
  (-1)^{v+1} \frac{N^v}{v!} \left.
	\left( t_1 \trace \frac{\delta}{\delta J} + t_2 \trace \frac{\delta^2}{\delta J^2} + \ldots \right)^v 
	e^{\frac{\lambda}{2N} \trace J^2}
	\right|_{J=0, \textrm{conn}}\, .
$$
A single monomial contribution in the parameters $t_d$ is
$$
  (-1)^{v+1} \frac{N^v}{S(\bfd)} \left.
	\prod_{i=1}^v t_{d_i} 
	\trace \left( \frac{\delta^{d_1}}{\delta J^{d_1}} \right) \ldots 
	\trace \left( \frac{\delta^{d_v}}{\delta J^{d_v}} \right) 
	e^{\frac{\lambda}{2N} \trace J^2}
	\right|_{J=0, \textrm{conn}}\, ,
$$
where the degrees $\bfd = (d_1,\ldots,d_v)$ are assumed to be in descending order, thus denoting a Young diagram. 
The symmetry factor $S(\bfd)$ is the order of the stabilizer group of $\bfd$ (as a subgroup of the symmetric group), e.g. for 
$\bfd = (7,7,7,4,4,3,3,3,3,3)$ the symmetry factor is $3!\cdot 2! \cdot 5!$.

As first observed in \cite{tHooft:1973alw}, the Feynman diagrams for large $N$ matrices can be related to ribbon graphs. 
Each vertex of the ribbon graph corresponds to a trace 
$\trace \left(\delta^{d_i}/\delta J^{d_i}\right)$, and each edge is weighted by $\lambda/N$. The vertex degrees are related to the edge number by 
$$
  2e = d_1 + \ldots + d_v\, , 
$$
in other words, the Young diagram $\bfd$ has even total degree $2e$.

A connected ribbon graph with $v$ vertices, $e$ edges, and $f$ faces is related to the genus $h$ of a connected, closed $2$-dimensional surfaces through the Euler formula
$$
	2-2h=v-e+f.
$$

The above monomial contribution thus has the form
$$
  (-1)^{v+1} N^{2-2h}\, \frac{ \lambda^e\, t_\bfd }{S(\bfd)}
	\, C_h(\bfd)	\, ,
$$
where $t_\bfd = t_{d_1} \cdot \ldots \cdot t_{d_v}$, and the numbers 
$C_h(\bfd)$ count the distinct Feynman diagrams corresponding to ribbon graphs of genus $h$ and degree $\bfd$. To obtain the correct counting, the vertices must be marked to account for distinct contributions for multiple vertices with the same degree. The vertices might simply be marked by natural numbers $1,\ldots,v$. Also, the half-edges of each vertex need to be decorated in cyclic order, say $1,\ldots,d_i$. (Equivalently, for oriented ribbon graphs, for each vertex a single half-edge is distinguished.) Following \cite{Mulase:2012yk}, where these combinatorial objects were studied in the context of matrix models and loop equations, let us denote the set of decorated marked ribbon graphs with $v$ vertices of degrees $(d_1,\ldots,d_v)$ by $\hat G_h(\bfd) = \hat G_h(d_1,\ldots,d_v)$. Then
$$
  C_h(\bfd) = \sum_{\Gamma \in \hat G_h(\bfd)} 1\, . 
$$
Since the simplest numbers $C_0(d)$ (for $d$ even) turn out to be the Catalan numbers, we adopt the notation of \cite{Mulase:2012yk} and refer to them as generalized Catalan numbers.

Putting everything together, $\cF = \cF_{\ni } + \hat \cF$, the free energy becomes
\begin{equation}
  \label{eq:free_energy}
  \cF(\alpha, \lambda, t_d, N) = \frac{N^2}{2} \ln\left(N/\lambda\right)
						  +  \sum_{h=0}^\infty \, N^{2-2h} 
							\sum_{\bfd \in Y_{\mathrm{ev}}}
							(-1)^{v+1} \frac{\lambda^e\, t_\bfd} {S(\bfd)}\, C_h(\bfd)\, ,
\end{equation}
where $Y_{\mathrm{ev}}$ is the set of Young diagrams of even total degree, that is $\sum_i d_i = 2e$.

The free energy $\cF$ of the Hermitian one-matrix model is therefore a generating function for the generalized Catalan numbers,
\begin{equation*}
  \left. \,
	\partial_{t_{d_1}} \ldots \partial_{t_{d_v}} \cF\, \right|_{t=0} =
	(-1)^{v+1} \lambda^e\, N^2 \, \cP_\bfd(N^{-2})\, ,
\end{equation*}
where the polynomial
\begin{equation}
  \label{eq:CatalanPolynomial}
	\cP_\bfd(N^{-2}) := \sum_{h=0}^{d(\bfd)} N^{-2h} C_h(\bfd)
\end{equation}
subsumes the generalized Catalan numbers for all genera for a given $\bfd$.
The maximum degree of the polynomial is generally 
$d(\bfd) = \left\lfloor \frac{1}{2} (e(\bfd)-v(\bfd) + 1) \right\rfloor$, which is determined by the Euler formula and the fact that the number of faces of a ribbon graph is positive. 

\subsection{Schwinger-Dyson equation and generalized Catalan numbers}

A full census of the generalized Catalan numbers was first given in \cite{Walsh:1972-1975} (in fact, decorated marked graphs were called dicings therein). The numbers $C_h(\bfd)$ of decorated marked graphs in $\hat G_h(\bfd)$ were shown to satisfy recursion relations,
\begin{eqnarray}
  \label{eq:genCatalanRecursion}
	C_h(d_1,\ldots, d_v) &=& \sum_{i=1}^{v-1} d_i \, C_h(d_1,\ldots, d_i + d_v -2, \ldots, d_{v-1}) \nonumber \\
	                     &+& \sum_{k=0}^{d_v-2} C_{h-1}(d_1,\ldots,d_{v-1},k,d_v-2-k) \nonumber \\
											 &+& \sum_{\stackrel{D_1 \cup D_2 = (d_1,\ldots,d_{v-1})}{D_1 \cap D_2 = \emptyset}}
											     \sum_{e+f=h} \sum_{k=0}^{d_v-2}
													 C_e(D_1,k) \, C_f(D_2,d_v-2-k)\, .
\end{eqnarray}
These relations fully determine the numbers starting from the initial value $C_0(0)=1$.

In the matrix model, the recursion relations (\ref{eq:genCatalanRecursion}) are a consequence of the loop equations \cite{Mulase:2012yk}. Here, we reproduce them as the Virasoro constraints that can be derived from the Schwinger-Dyson equation for degree $n \geq 0$:
\begin{equation}
	\label{eq:SchwingerDysonD0}
  \int DM \sum_{i,j} 
	\frac{\partial}{\partial M_{ij}} \left(M^{n+1}{}_{ij} \, e^{-S}\right) = 0\, .
\end{equation}
Applying the differential under the integral gives insertions of single trace operators $\trace{M^d}$, which lead to the Virasoro constraints
\begin{equation}
  \label{eq:VirasoroConstraint}
  -\lambda^{-1} \partial_{t_{n+2}} Z = N^2 \delta_{n,0} Z +
	    \sum_d d \, t_d \, \partial_{t_{d+n}} Z -
			2 \, \partial_{t_n} Z + 
			N^{-2} \hspace{-5pt}
			\sum_{\stackrel{k,l\geq 1}{k+l=n}} 
			\hspace{-5pt}
			\partial_{t_k}\,\partial_{t_l}\, Z\, .
\end{equation}
Notice also that 
$2\lambda \partial_\lambda Z = -\lambda^{-1} \partial_{t_2} Z$.

Now using the expression (\ref{eq:free_energy}) for the free energy as the generating function of the generalized Catalan numbers, recursion relations for the polynomials $\cP_\bfd=\cP_\bfd(N^{-2})$ can be derived order by order in $\lambda$. Setting $d_v=n+2$, we obtain with the abbreviation $\bfd_{v-1} = (d_1,\ldots,d_{v-1})$:
\begin{eqnarray}
  \label{eq:recursive_polynomial_eq}
  \cP_{(\bfd_{v-1},d_v)} &=& 
	\sum_{i=1}^{v-1} d_i \, 
	\cP_{(d_1,\ldots,d_i+d_v-2,\ldots,d_{v-1})} + \\
	&+& \hspace{-5pt} \sum_{\stackrel{k,l\geq 0}{k+l=d_v-2}} 
	\sum_{\stackrel{\bfd_1 \cup \bfd_2 = \bfd_{v-1}}
	{\bfd_1 \cap \bfd_2 = \emptyset}} 
	\hspace{-5pt}
	\cP_{(\bfd_1, k)} \, \cP_{(\bfd_2,l)} \, +\,  
	N^{-2} \hspace{-5pt}\sum_{\stackrel{k,l\geq 1}{k+l=d_v-2}} 
	\cP_{(\bfd_{v-1},k,l)} \, . \nonumber
\end{eqnarray}
Here, we formally introduced $\cP_{(0)} = 1$ and $\cP_{\bfd} = 0$, if at least one of the degrees in $\bfd$ is zero. The initial relation coming from the first term of the right-hand side of (\ref{eq:VirasoroConstraint}) is $\cP_{(2)} = \cP_{(0)}\cP_{(0)} = 1$. 

Using (\ref{eq:CatalanPolynomial}), it is straight forward to see that these polynomial relations are equivalent to the recursion relations (\ref{eq:genCatalanRecursion}) for the generalized Catalan numbers of \cite{Walsh:1972-1975,Mulase:2012yk}. In the next section we will see that a refinement of these polynomials will play a key role in higher dimensions.

\section{The matrix model in arbitrary dimensions}
\label{sec:MMAD}

Let us work out the perturbative expansion of the matrix model for $D>0$ defined in (\ref{eq:higherdim_matrix_model}).
As anticipated in the introduction the Greens function for the non-local kinetic term $e^{-\frac{\alpha}{2}\Delta}$ is the heat kernel $G(\alpha,x,x')$ for the associated heat equation
$$
  \partial_\alpha G(\alpha,x,x') = \frac{1}{2} \Delta_x \, G(\alpha,x,x')\, .
$$
On $\cM = \bbR^D$ with Euclidean metric, the non-local kinetic term in (\ref{eq:higherdim_matrix_model}) requires the boundary condition that the matrix field $M(x)$ and all its derivatives must vanish at infinity. This is reflected in the Greens function given by the Gaussian distribution function
\begin{equation*}
  G(\alpha,x,x') = \frac{1}{(2\pi\alpha)^{D/2}} 
	e^{-\frac{(x-x')^2}{2\alpha}}\, .
\end{equation*}

When applying Feynman rules to obtain the free energy from all connected Feynman diagrams, the graph combinatorics is governed by decorated marked ribbon graphs, just as for the Hermitian one-matrix model. The sole difference for $D>0$ is the fact that each connected ribbon graph is weighted by a product of Gaussian distributions for all edges and an integration over $\cM$ for each vertex. Collecting all contributions the perturbation expansion gives
\begin{equation}
  \hat \cF(\alpha, \lambda, t_d, N) = 
	\sum_{h=0}^\infty N^{2-2h} 
	\sum_{\bfd \in Y_{\mathrm{ev}}} \; {(-1)^{v+1}
	\frac{\lambda^{e}t_\bfd}{S(\bfd)} \hspace{-5pt} 
	\sum_{\Gamma\in \hat G_h(\bfd)} \hspace{-5pt} 
	\int d^{Dv}X
	\frac{1}{\left(2\pi \alpha\right)^{D v/2}} \hspace{-5pt} 
	\prod_{e_{ij}\in E(\Gamma)} \hspace{-5pt} 
	e^{-\frac{(x_i-x_j)^2}{2\alpha}}}.
  \label{eq:W_eff_perturbative}
\end{equation}
Here $X=(x_1,\ldots,x_v)$, and $E(\Gamma)$ denotes the set of edges in the decorated marked ribbon graph $\Gamma$, and $e_{ij}$ is an edge connecting vertices $i$ and $j$.
The expression (\ref{eq:W_eff_perturbative}) explicitly shows the relation of the matrix model to the discrete string perturbation expansion with string coupling $N^{-1}$. The summation over Young diagrams and ribbon graphs for fixed genus $h$ corresponds to the integration over world sheet metrics in the continuum description.

In the discrete setting the embedding coordinates $X$ can be integrated out almost completely by rewriting the product of heat kernels for the ribbon graph $\Gamma$ in terms of its Laplacian matrix. 

\begin{figure*}
  \centering
	\begin{tabular}{ccc}
		\includegraphics[width=0.2\textwidth]{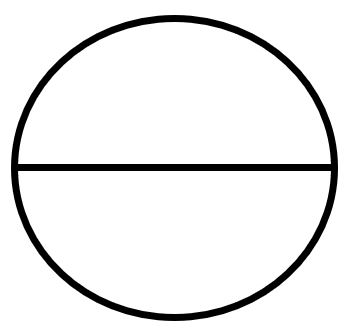} &
		\includegraphics[width=0.2\textwidth]{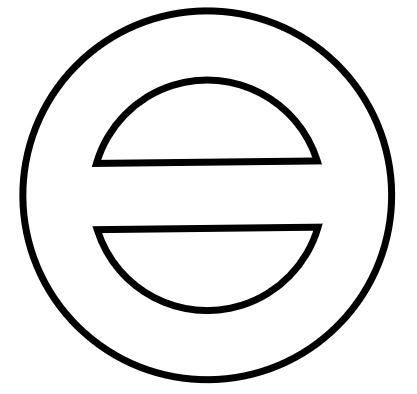} &
		\includegraphics[width=0.19\textwidth]{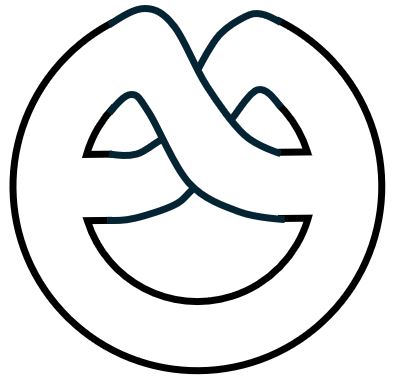} \\
		skeleton & genus 0 & genus 1 
	\end{tabular}
	\caption{Example of ribbon graphs giving rise to the same skeleton}
	\label{fig:RibbonGraphsSkeleton}
\end{figure*}

Let us introduce some relevant definitions and notation for the Laplacian matrix of a graph and its adjacency matrix. Both matrices depend on the skeleton $\gamma$ of the ribbon graph $\Gamma$ only, where we adopted the convention of \cite{Bessis:1980ss} and use the notion skeleton for a graph $\gamma$ that arises from a ribbon graph $\Gamma$ by forgetting the cyclic ordering of the half-edges, and dropping the decoration of half-edges and marking of vertices. Figure \ref{fig:RibbonGraphsSkeleton} shows an example. 
Let us denote the map from the set of decorated marked ribbon graphs $\hat G_h(\bfd)$ to the set of (unmarked) skeletons $Sk(\bfd)$ by
\begin{eqnarray}
	\label{eq:RibbonSkeletonMap}
	\operatorname{sh}_h &:& \hat G_h(\bfd) \rightarrow Sk(\bfd), \\
	                  && \Gamma \mapsto \gamma. \nonumber
\end{eqnarray}
Note that the map preserves the number of vertices, but forgets the genus $h$. 

The adjacency matrix $A(\gamma)$ is the symmetric matrix filled with the number of edges between vertices $i$ and $j$ at the position $(i,j)$. If the skeleton has loops at some vertices, the adjacency matrix bears twice the number of loops at each vertex on the diagonal. Each row or column adds up to the corresponding vertex multiplicity, $\bfd= (d_1,\ldots,d_v)$, which can be collected in a diagonal matrix $D(\gamma)$. The skeleton $\gamma$ is uniquely determined by the adjacency matrix $A(\gamma)$ (up to similarity transformations that permute rows and columns). 

The Laplacian matrix for a skeleton $\gamma$ then is
$$
	\Delta(\gamma) = D(\gamma) - A(\gamma).
$$
Each row or column of the Laplacian matrix adds up to zero and in fact the rank of the Laplacian matrix for a connected skeleton is $v-1$. The determinant of the Laplacian matrix vanishes, but the determinant of the reduced Laplacian matrix $\Delta'(\gamma)$ (removing some row $i$ and column $i$ for arbitrary $i$) is a graph invariant that plays an important role in graph theory as a measure of graph complexity, and will prominently show up in the following as well.

Coming back to the free energy (\ref{eq:W_eff_perturbative}), the exponential weight factors can be rewritten in terms of the Laplacian matrix:
$$
	\sum_{e_{ij}\in E(\gamma)} (x_i-x_j)^2 = X \Delta(\gamma) X\, .
$$
Using the property that the Laplacian matrix $\Delta(\gamma)$ of a connected skeleton $\gamma$ has rank $v-1$, we can perform a base change $X \rightarrow (X',x)$ with $X'=(x_1,\ldots,x_{v-1})$ to remove the last column and row from $\Delta(\gamma)$, thus ending up with the reduced Laplacian $\Delta'(\gamma)$. To be precise, for the base change we select the coordinate of one of the vertices of the skeleton $\gamma$ and use it as reference, say $x = x_v$. The other coordinates are shifted with respect to the reference, $X=(x_1,\ldots,x_v) \mapsto (X',x)=(x_1-x_v,\ldots,x_{v-1}-x_v, x_v)$. As a result the last row and column of the Laplacian matrix is set to zero, that is
$$
  X \Delta(\gamma) X = X' \Delta'(\gamma) X'\, .
$$
The free energy then becomes
\begin{equation}
	\label{eq:free_energy_gaussian}
  \hat \cF(\alpha, \lambda, t_d, N) = 
	\sum_{h=0}^\infty N^{2-2h} 
	\sum_{\bfd \in Y_{\mathrm{ev}}} \; {(-1)^{v+1}
	\frac{\lambda^{e}\, t_\bfd}{S(\bfd)} \hspace{-2pt} 
	\sum_{\Gamma\in \hat G_h(\bfd)} \hspace{-2pt} 
	\int d^Dx \int d^DX' \frac{1}{\left(2\pi \alpha\right)^{Dv/2}} 
	e^{-\frac{X'\Delta'(\gamma)X'}{2\alpha}}}\, ,
\end{equation}
where $d^DX' = \prod_{i=1}^{v-1}d^Dx_i$. The Gaussian distribution for $X'$ is known from the theory of Gaussian molecules \cite{Eichinger:1980}, a connection that will be exploited in Section \ref{sec:Gyration}. In fact, the free energy can be interpreted as generating function of Gaussian molecules in $D$ dimensions.

The integral over $X'$ can readily be computed and results in
\begin{equation*}
  \hat \cF(\alpha, \lambda, t_d, N) = 
  \frac{V_\cM}{(2\pi\alpha)^{D/2}} 
  \sum_{h=0}^\infty N^{2-2h} 
	\sum_{\bfd \in Y_{\mathrm{ev}}} \; {(-1)^{v+1}
	\frac{\lambda^{e}\, t_\bfd}{S(\bfd)} \hspace{-2pt} 
	\sum_{\Gamma\in \hat G_h(\bfd)} \frac{1}{(\det \Delta'(\gamma))^{D/2}}}\, .
\end{equation*}
We find that in the free energy the sum over decorated marked ribbon graphs is weighted by the factor $(\det \Delta'(\gamma))^{-D/2}$ \cite{Kazakov:1988ch}, and the free energy is proportional to the infinite space volume $V_\cM = \int d^Dx$.

Let us make contact with the enumeration of decorated marked ribbon graphs as reviewed in the previous section.
Notice that considering the map (\ref{eq:RibbonSkeletonMap}) from ribbon graphs to skeletons, the preimage $\operatorname{sh}_h^{-1}(\gamma) \subset \hat G_h(\bfd)$ contains all decorated marked ribbon graphs for the skeleton $\gamma$.
The number of marked ribbon graphs of genus $h$ that map to the skeleton $\gamma \in Sk(\bfd)$ are counted by
$$
	C_{h}(\bfd, \gamma) := \sum_{\Gamma\in \operatorname{sh}_h^{-1}(\gamma)}\hspace{-5pt} 1\, .
$$
We therefore obtain
$$
  \sum_{\Gamma\in \hat G_h(\bfd)} 
	\frac{1}{(\det \Delta'(\gamma))^{D/2}}\, =
	\sum_{\gamma \in Sk(\bfd)}\,
	\frac{C_{h}(\bfd, \gamma)}{(\det \Delta'(\gamma))^{D/2}}
$$

For a given skeleton $\gamma \in Sk(\bfd)$, the numbers $C_h(\bfd,\gamma)$ can be summarized in an invariant graph polynomial, 
\begin{equation}
  \cP_{\mathbf{d}, \gamma}(N^{-2}) := 
	\sum_{h=0}^{d(\bfd)} 
	N^{-2h}\, C_{h}(\mathbf{d}, \gamma),
  \label{eq:graph_polynomial}
\end{equation}
Table \ref{tab:invariant_polynomials} shows some examples for invariant polynomials $\cP_{\bfd, \gamma}(N^{-2})$,
which were determined by explicit counting of Feynman diagrams, that is, decorated marked ribbon graphs. 
\begin{table}[!ht]
	\centering
	\resizebox{0.90\textwidth}{!}{%
	\begin{tabular}{|cccc|}
		\hline
		$\bfd=(2,2),\quad \det\Delta'=2$ & $(3,3),\quad 3$ & $(3,3),\quad 1$ & $(4,2),\quad 2$ \\
		\multicolumn{4}{|c|}{\includegraphics[width=1.1\textwidth]{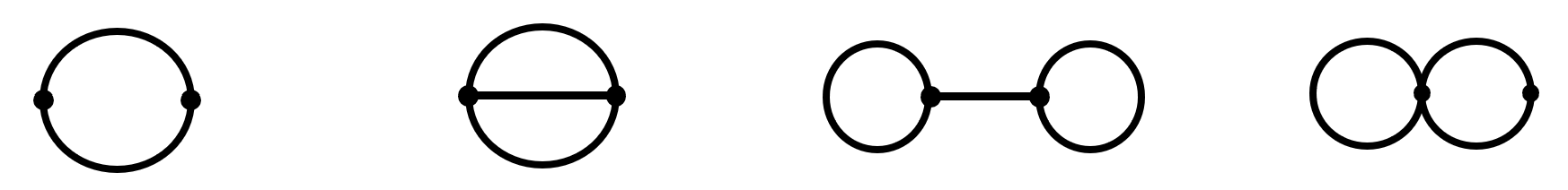}}  \\
		$\cP_{\bfd,\gamma}(N^{-2})=2$ & $3+3 N^{-2}$ & $9$ & $8+4 N^{-2}$ \\
		\hline
		$(4,4),\quad 4$ & $(4,4),\quad 2$ & $(5,3),\quad 3$ & $(5,3),\quad 1$ \\
		\multicolumn{4}{|c|}{\includegraphics[width=1.1\textwidth]{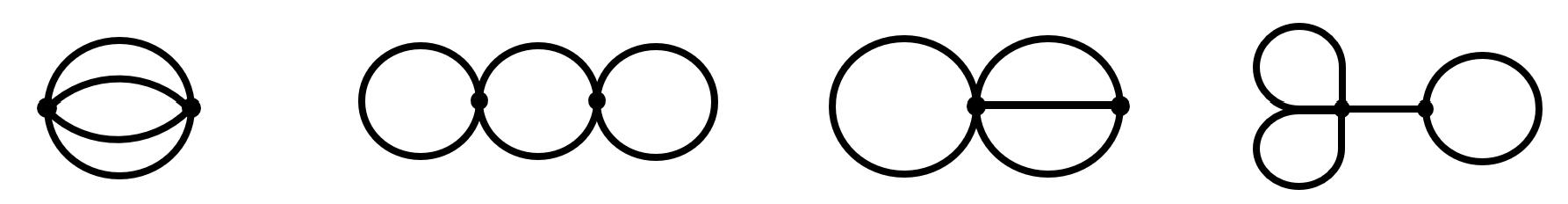}}  \\
		$4+20 N^{-2}$ & $32+40 N^{-2}$ & $15+45 N^{-2}$ & $30+15 N^{-2}$ \\
		\hline
		$(5,5),\quad 5$ & $(5,5),\quad 3$ & $(5,5),\quad 1$ & $(6,2),\quad 2$ \\
		\multicolumn{4}{|c|}{\includegraphics[width=1.1\textwidth]{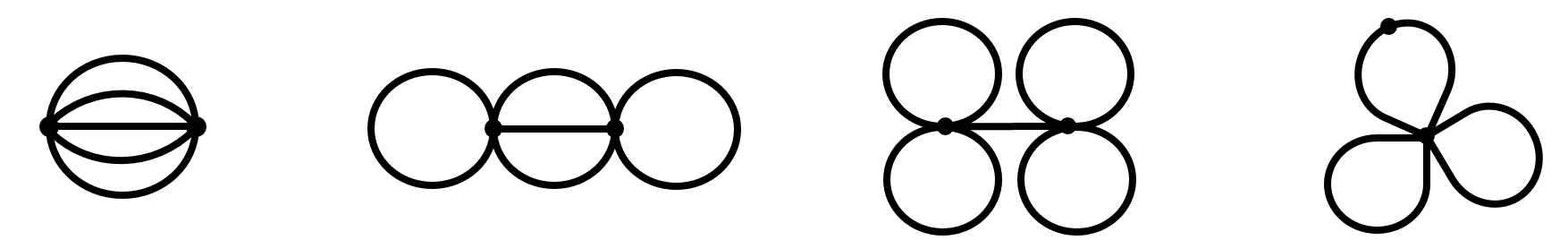}}  \\
		$5+75 N^{-2}+40 N^{-4}$ & $75+425 N^{-2}+100 N^{-4}$ &
		$100+100 N^{-2}+25 N^{-4}$ & $30+60 N^{-2}$ \\
		\hline
		$(3,3,2),\quad 5$ & $(3,3,2),\quad 2$ & $(3,3,2),\quad 1$ & $(4,2,2),\quad 4$ \\
		\multicolumn{4}{|c|}{\includegraphics[width=1.1\textwidth]{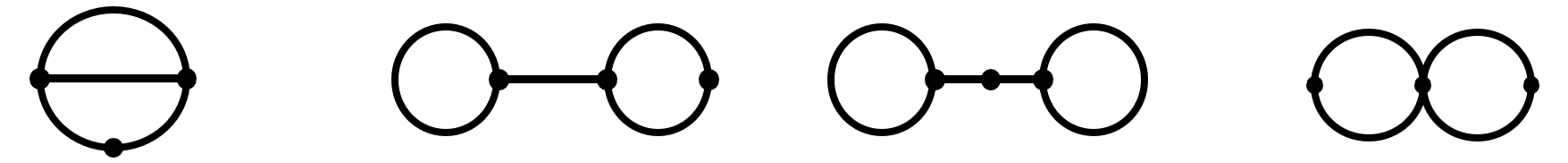}}  \\
		$18+18 N^{-2}$ & $36$ & $18$ & $16+8N^{-2}$ \\
		\hline
		$(4,2,2),\quad 3$ & $(4,4,2),\quad 7$ & $(4,4,2),\quad 3$ & $(4,4,2),\quad 4$ \\
		\multicolumn{4}{|c|}{\includegraphics[width=1.1\textwidth]{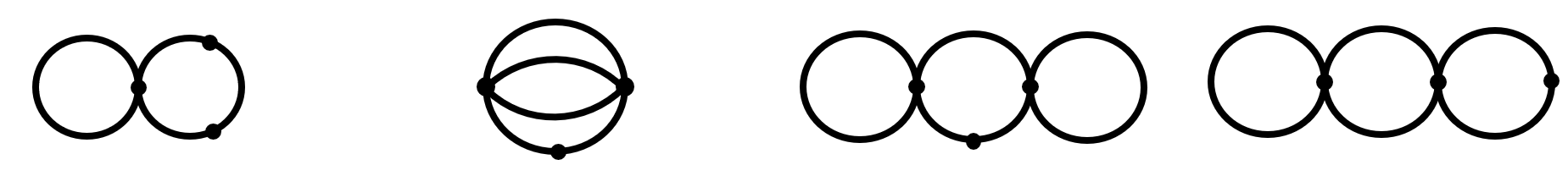}}  \\
		$32+16 N^{-2}$ & $32+160 N^{-2}$ & $128+160 N^{-2}$ & $128+160 N^{-2}$ \\
		\hline
		$(4,4,4),\quad 12$ & $(4,4,4),\quad 7$ & $(4,4,4),\quad 4$ & $(4,4,4),\quad 3$ \\
		\multicolumn{4}{|c|}{\includegraphics[width=1.1\textwidth]{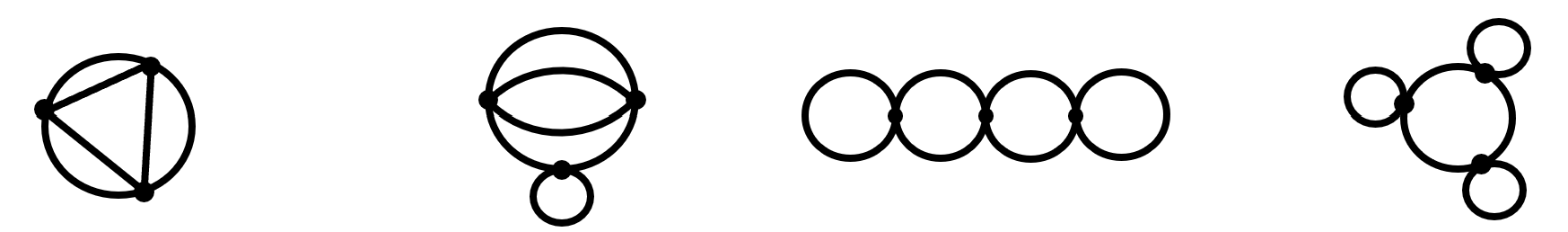}}  \\
		$64+1088 N^{-2}+576 N^{-4}$ & $384+2496 N^{-2}+576 N^{-4}$ &
		$768+1536 N^{-2}+288 N^{-4}$ & $512+1216 N^{-2}$ \\
		\hline
		$(3,3,3,3),\quad 16$ & $(3,3,3,3),\quad 12$ & $(3,3,3,3),\quad 5$ & $(3,3,3,3),\, 2$ \qquad $(3,3,3,3),\, 1$ \\
		\multicolumn{4}{|c|}{\includegraphics[width=1.1\textwidth]{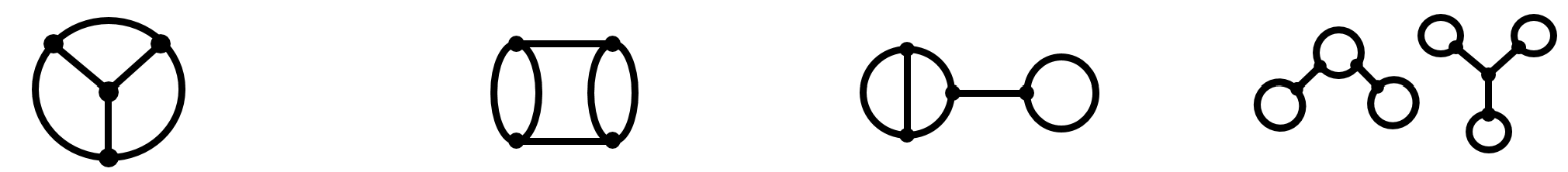}}  \\
		$162+1134 N^{-2}$ & $486+1458 N^{-2}$ &
		$1944 + 1944 N^{-2}$ & $1944$ \qquad $648$\\
		\hline
	\end{tabular}}
	\caption{Various skeletons $\gamma$ and their invariant polynomials $\cP_{\mathbf{d}, \gamma}(N^{-2})$, 
	and the determinants of their reduced Laplacian matrix.}
	\label{tab:invariant_polynomials}
\end{table}

Expressing the free energy in terms of the invariant polynomials 
$\cP_{\bfd, \gamma}(N^{-2})$ gives
\begin{equation}
	\label{eq:free_energy_w}
  \hat \cF(\alpha,\lambda,t_d,N) = 
	\frac{V_\cM}{(2\pi\alpha)^{D/2}} \, \hat w(\lambda, t_d, N)
\end{equation}
with 
\begin{equation}
	\label{eq:FreeEnergyInvariantPolynomials}
	\hat w(\lambda, t_d, N) = 
	\sum_{\mathbf{d} \in Y_{\textit{ev}}} (-1)^{v+1}
	\frac{\lambda^e\, t_{\mathbf{d}}}{S(\mathbf{d})} 
	\sum_{\gamma\in Sk(\bfd)} 
	\frac{N^2\, \cP_{\mathbf{d}, \gamma}(N^{-2})}{\det\Delta'(\gamma)^{D/2}}.
\end{equation}
Notice that all $\alpha$ dependence of $\hat \cF$ comes with the volume factor. $\hat w(\lambda,t_d,N)$ does only depend on the dimensionless coupling constants $\lambda$ and $t_d$ as well as on $N$ and $D$. The dependence on $D$ is simple, it controls the weight of the determinant of the reduced Laplacian matrix $\det\Delta'(\gamma)$. The next subsection will elaborate on this fact. For later reference we introduce the expectation value of a graph invariant, say $\hat \cO(\gamma)$, with respect to $\hat w$ by
$$
  \langle \hat \cO(\gamma) \rangle_{\hat w} := 
    \frac{1}{\hat w} \sum_{\mathbf{d} \in Y_{\textit{ev}}} (-1)^{v+1}
	\frac{\lambda^e\, t_{\mathbf{d}}}{S(\mathbf{d})} 
	\sum_{\gamma\in Sk(\bfd)} \hat \cO(\gamma) \,
	\frac{N^2\, \cP_{\mathbf{d}, \gamma}(N^{-2})}{\det\Delta'(\gamma)^{D/2}}.
$$

The coefficients $C_{h}(\mathbf{d}, \gamma)$ of the invariant graph polynomials are a refinement of the generalized Catalan numbers $C_{h}(\mathbf{d})$ of the previous section. The special case $D=0$ shows that summing the numbers $C_h(\bfd,\gamma)$ for all skeletons $\gamma$ that share the same degrees $\bfd$ and genus $h$ must reproduce the generalized Catalan numbers,
\begin{equation}
	C_{h}(\mathbf{d}) = \sum_{\gamma \in Sk(\bfd)} C_{h}(\bfd,\gamma) .
  \label{eq:relation_to_generalized_catalan}
\end{equation}
In view of this relation, the enumerations of Table \ref{tab:invariant_polynomials} can be compared with some results on generalized Catalan numbers in \cite{GopalaKrishna:2018wzq} or with solutions of the recursion relations (\ref{eq:genCatalanRecursion}) or (\ref{eq:recursive_polynomial_eq}). 
It is not clear to the author whether the graph polynomials $\cP_{\bfd,\gamma}(N^{-2})$ as introduced in (\ref{eq:graph_polynomial}) were studied before in the literature or whether they are related to known invariant graph polynomials. 
It would be interesting to see whether the refined generalized Catalan numbers $C_h(\bfd,\gamma)$, or equivalently the invariant polynomials $\cP_{\mathbf{d}, \gamma}(N^{-2})$, satisfy some refined version of the recursion relations (\ref{eq:genCatalanRecursion}).

Taking the sum over skeletons $\gamma$, we can consider the polynomials 
$$
  \cP^D_\bfd(N^{-2}) := \sum_{\gamma\in Sk(\bfd)} 
	\frac{\cP_{\mathbf{d}, \gamma}(N^{-2})}{\det\Delta'(\gamma)^{D/2}} =
	\sum_h N^{-2h} \sum_{\gamma\in Sk(\bfd)} 
	\frac{C_h(\bfd,\gamma)}{\det\Delta'(\gamma)^{D/2}}
	\, ,
$$
which are the immediate generalization of the polynomials (\ref{eq:CatalanPolynomial}): $\hat w$ has exactly the same structure as the free energy $\hat \cF$ in (\ref{eq:free_energy}), but with $\cP_\bfd(N^{-2})$ replaced by $\cP_\bfd^D(N^{-2})$.
Notice that for even dimension $D$ the polynomials $\cP^D_\bfd(N^{-2})$ have rational coefficients, and only for $D=0$ the coefficients are non-negative integers. 

Although a recursive procedure to compute the polynomials $\cP_{\mathbf{d}, \gamma}(N^{-2})$ is not known to the author, we can still state some properties for the simplest polynomials. 

A skeleton $\gamma$ with a single vertex of degree $\bfd = (d)$ is unique. Its Laplacian matrix is a $1\times 1$ zero-matrix, and its reduced determinant in the free energy (\ref{eq:FreeEnergyInvariantPolynomials}) is formally given by $\det \Delta'(\gamma) = 1$. The graph polynomials counting ribbon graphs with a single vertex are therefore given by
$$
  \cP^D_{d, \gamma}(N^{-2}) = \cP_{d,\gamma}(N^{-2}) = 
	\cP_{d}(N^{-2}),
	\quad \mathrm{for~all}~d \in \bbN_{\mathrm{ev}}\, .
$$
In particular, the polynomials for skeletons with a single vertex do not depend on the dimension $D$, and the refined Catalan numbers are just the generalized Catalan numbers, $C_h(d, \gamma) = C_h(d)$ for all $h \geq 0$.

\subsection{Schwinger-Dyson equation}
\label{sec:SchwingerDyson}

Just as for $D=0$ we could ask whether for $D>0$ the Schwinger-Dyson equations lead to Virasoro constraints that completely determine the graph polynomials. There are several ways to generalize equation (\ref{eq:SchwingerDysonD0}) in the higher-dimensional context. Let us consider the equation
$$
  \int DM \sum_{i,j} 
	\int d^D x \frac{\partial}{\partial M_{ij}(x)} \left(M^{n+1}(x)_{ij} \,
	e^{-S}\right) = 0\, .
$$
By applying the derivative with respect to the matrix field, we immediately realize that contact terms $\delta^D(0)$ as well as double-trace operators $\int d^Dx \trace{M(x)^k} \trace{M(x)^l}$ will appear. The latter prevent us from applying the procedure known from matrix models to arrive at Virasoro constraints. One resolution could be to promote the parameters $t_d$ to slowly varying background fields $t_d(x)$, so that the double-trace operators can be reproduced by the functional derivative $\partial_{t_k(x)}\partial_{t_l(x)} Z$. But this comes at the price of introducing  even more contact terms. 

We do not follow this route here, but stay with $t_d$ as parameters and merely consider the simplest Schwinger-Dyson equation for $n=0$. It can be written in terms of the partition function $Z = Z_\ni \cdot \hat Z$ as
$$
  2\lambda \partial_\lambda Z = N^2 V_\cM\, \delta^D(0) Z + \sum_d d\, t_d\,  \partial_{t_d} \, Z\, .
$$
Notice that setting $t_d = 0$ for all $d\geq 1$ gives
$$
  2\lambda \partial_\lambda Z_\ni = N^2 V_\cM \delta^D(0) Z_\ni \, , 
$$
which implies that
$$
  Z_\ni = C(\alpha,N,D)\, \lambda^{\frac{N^2}{2} V_\cM \delta^D(0)} \, .
$$
This is in line with the result (\ref{eq:NonIntPartition}) for $Z_\ni$ in Appendix \ref{apdx:free_partition_function}.

Expressing the Schwinger-Dyson equation for $n=0$ in terms of the free energy $\hat\cF = - \ln(Z/Z_\ni) = \frac{V_\cM}{(2\pi\alpha)^{D/2}} \hat w$, we obtain
$$
  2\lambda\partial_\lambda \hat w = 
	\sum_d d\, t_d\, \partial_{t_d}\, \hat w\, .
$$
From the perturbation expansion (\ref{eq:FreeEnergyInvariantPolynomials}) of the free energy, it can be readily seen that the Schwinger-Dyson equation for $n=0$ is simply a consequence of the graph combinatorial property $2e = \sum_i d_i$.

\subsection{Spanning tree entropy and graph complexity}
\label{sec:GraphComplexity}

Kirchhoff's matrix tree theorem is a well-known result in graph theory. It states that for a connected graph $\gamma$, the number of spanning trees $\tau(\gamma)$ can be computed by the determinant of the reduced Laplacian matrix,
$$
  \tau(\gamma) = \det\Delta'(\gamma)\ .
$$
The number of spanning trees is a measure for the complexity of a graph. In the mathematics literature, efforts were undertaken to estimate the graph complexity in the limit of a large number of vertices, work going back to \cite{McKay:1981-1983}. In \cite{Lyons:2005} the spanning tree entropy
$$
  \bfh(\gamma) = \frac{1}{v} \ln{\tau(\gamma)},
$$
was used to study the limiting behavior for large vertex number $v \rightarrow \infty$.
It measures the exponential dependence of $\tau(\gamma)$ on $v = v(\gamma)$. 
By abuse of notation we directly refer to $\bfh(\gamma)$ as spanning tree entropy. 

\begin{figure*}[!t]
  \centering
	\includegraphics[width=0.6\textwidth]{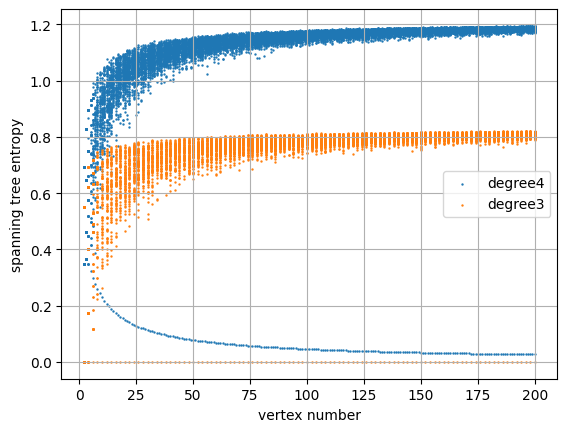}
	\caption{Spanning tree entropy $h(\gamma)$ over vertex number for a random sample of skeleton graphs. The value $h(\gamma) = 0$ corresponds to tree graphs (here with loops at the ends of the tree). The "gap" between the upper and lower lying values is just an artefact of drawing the sample. Values in the gap are sparse and unlikely. For large $v$ the value for a randomly chosen graph approaches $h_d$, here $h_3=\ln(2^2/\sqrt{3}) \approx 0.837$ and $h_4=\ln(3^3/2^3) \approx 1.216$.} 
	\label{fig:spanning_tree_entropy}
\end{figure*}

For $d$-regular connected graphs, generated by a monomial potential $V(M)$ of degree $d$, the limiting behavior for large $v$ was found in \cite{McKay:1981-1983,Lyons:2005} to be
\begin{equation}
	\label{eq:entropy_limit}
  \bfh(\gamma) \rightarrow h_d :=
	\ln{\left(\frac{(d-1)^{d-1}}{(d^2-2d)^{d/2-1}}\right)}\, .
\end{equation}
This is reflected in Figure \ref{fig:spanning_tree_entropy}, which shows the spanning tree entropy for random samples of connected degree $3$ and $4$ regular graphs.
In \cite{McKay:1981-1983,Greenhill:2014} it was shown that the limiting behavior of the number of spanning trees $\tau(\gamma)$ has in fact exponential-like growth $\cO(v^{-1} e^{h_d v})$ for $d\geq 3$. 

In the free energy (\ref{eq:free_energy_w}) the spanning tree entropy appears through the reduced Laplacian determinant in $\hat w(\lambda, t_d, N)$.
Analytically continuing $D$ and differentiating with respect to the dimension we obtain
$$
	\partial_D \ln{\hat w} =
	-\frac{1}{2} \left< \ln{\hat\tau} \right>_{\hat w} =
	-\frac{1}{2} \left< \hat v\, \hat\bfh \right>_{\hat w}
	\longrightarrow 
	-\frac{h_d}{2} \left< \hat v\right>_{\hat w}
	\ .
$$
The dependence of the matrix model free energy on the dimension $D$ is therefore controlled by the spanning tree entropy.

\subsection{Radius of gyration for Gaussian molecules}
\label{sec:Gyration}

In this section we intend to estimate the size of the embedding of a graph $\gamma$ in $\cM$. To this end, we notice that the Gaussian distribution in (\ref{eq:free_energy_gaussian}) is the defining probability distribution for Gaussian molecules \cite{Eichinger:1980}, which consider molecules as graph embeddings into $\cM = \bbR^D$ that are subject to Gaussian interactions along the edges of the graph:\footnote{Different from the mathematics literature where the distribution is given by the standard normal distribution per edge we consider variance $\alpha$. Also note that we fixed translational invariance by using the coordinates $X = (X',0)$.}
$$
  \langle~ f(X)~ \rangle_\gamma = \frac{(\det \Delta')^{D/2}}{\left(2\pi \alpha\right)^{D(v-1)/2}} 
	\int d^DX' {e^{-\frac{X'\Delta'(\gamma)X'}{2\alpha}} f(X)} \, .
$$

The size of a Gaussian molecule in $\cM$ can be measured by the expectation value of the gyration radius
$$
  R^2_\gyr(X) := \frac{1}{v} \sum_{i=1}^v (x_i - \bar x)^2 = \frac{1}{2v^2} \sum_{i,j=1}^v (x_i - x_j)^2\, .
$$

For a given graph $\gamma$ we define its size by
$$
  R^2_\gyr(\gamma) := \langle R^2_\gyr(X) \rangle_\gamma = \alpha D
  \left(
	\frac{1}{v} \sum_{i=1}^{v-1} \Delta'{}_{ii}^{-1} - 
	\frac{1}{v^2} \sum_{i,j=1}^{v-1} \Delta'{}_{ij}^{-1}
  \right)\, .
$$
\begin{figure*}[!t]
  \centering
    \includegraphics[width=0.66\textwidth]{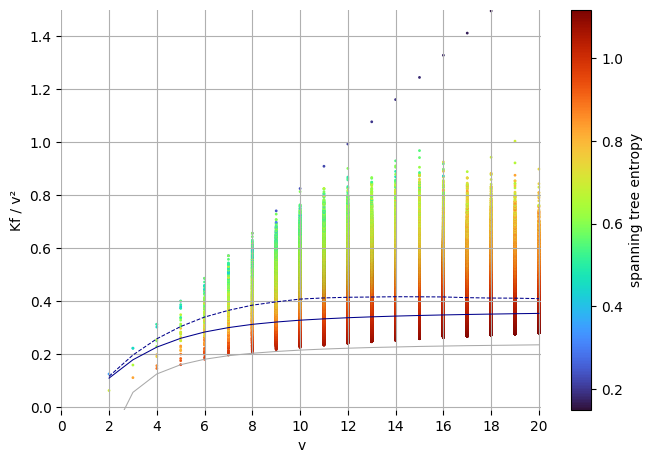}
	\caption{For a random sample of regular graphs of degree $4$, $r^2_\gyr(\gamma)=\kh(\gamma)/v^2$ is plotted over the vertex number $v$. The lower bound of (\ref{eq:Kirchhoff_bounds}) is indicated as gray line. The average value for $D=0$ (full blue line) approaches the large $v$ value $S_4(4) = 3/8$ derived from the Kesten-McKay distribution. The dashed line shows the average for $D=2$, that is weighted by $\det\Delta'^{-D/2}$. The color coding depicts the spanning tree entropy $h(\gamma)$. Highly connected graphs have low gyration radius, whereas cyclic graphs have large gyration radius.}
	\label{fig:gyration_radius}
\end{figure*}
A well-known result from the study of Gaussian molecules \cite{Eichinger:1980,Cantarella:2022} relates it to the Kirchhoff index,
\begin{equation*}
  	R^2_\gyr(\gamma) = \alpha D\, \frac{\kh(\gamma)}{v^2} \, .
\end{equation*}
The Kirchhoff index is a measure for the connectivity of a graph $\gamma$. It was introduced in \cite{Klein:1993ri} as the sum of resistances $r_{ij}$ over all pairs of vertices $i$ and $j$, where each edge has assigned unit resistance. 
The Kirchhoff index can be computed in terms of the Laplacian matrix $\Delta(\gamma)$ and is given by the sum of non-vanishing Eigenvalues $\lambda_i$ of $\Delta$, see \cite{Gutman:1996qwki} and references therein:
$$
  \kh(\gamma) := \sum_{i<j} r_{ij} = v \sum_{i=2}^{v} 1/\lambda_i\, .
$$
For what follows we introduce
$$
  r^2_\gyr(\gamma) := \frac{\kh(\gamma)}{v^2} \, ,
$$
and by abuse of notation also call it gyration radius.

\begin{figure*}[!t]
  \centering
    \includegraphics[width=0.3\textwidth]{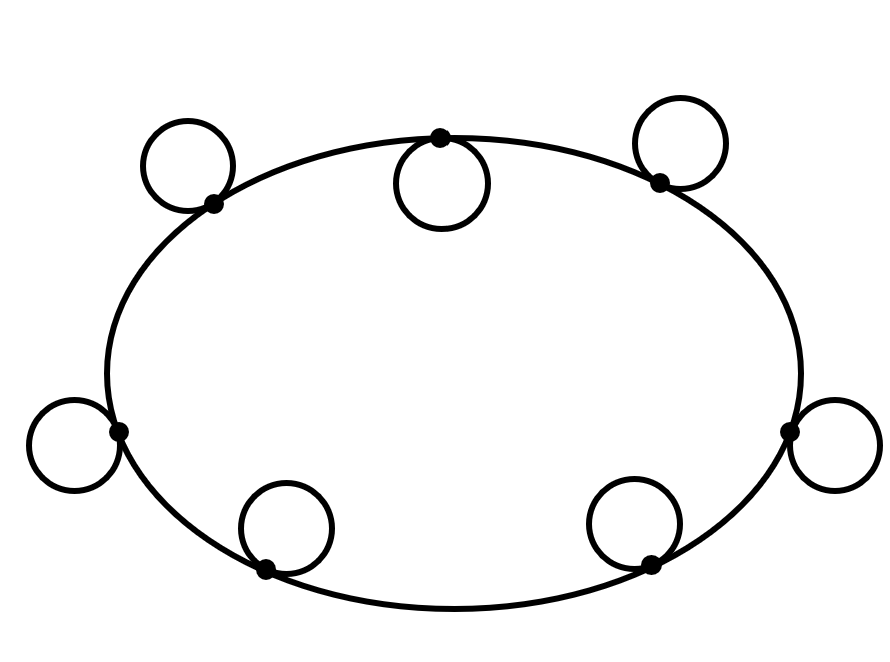}
	\caption{Cycle graph of degree $4$ with loops at each vertex.}
	\label{fig:long_chains}
\end{figure*}

For $d$-regular graphs with all vertices having the same degree, lower and upper bounds for the Kirchhoff index are known \cite{Palacios:2016}, here expressed in terms of the gyration radius $r^2_\gyr(\gamma)$:
\begin{equation}
	\label{eq:Kirchhoff_bounds}
	\frac{\left( v - 1 \right)}{dv} - \frac{1}{v^2} ~\leq~r^2_\gyr(\gamma) ~\leq~ 
  	\left\{
		\begin{array}{cl}
			\frac{(v^2-1)}{12v}  & \textrm{for}~d~\textrm{even}, \\[6pt]
			\frac{(v^2-1)}{6v} & \textrm{for}~d~\textrm{odd}.
		\end{array}
	\right.
\end{equation}
These bounds are refleced in the random sample shown in Figure \ref{fig:gyration_radius}. 
Loosly connected graphs have a large Kirchhoff index at the upper end of the distribution. The upper bound in (\ref{eq:Kirchhoff_bounds}) is saturated by tree graphs (with loops at the end) for odd $d$ \cite{Palacios:2016}, and saturated by cycle graphs (see Figure \ref{fig:long_chains}) for even $d$ \cite{Zhang:2007,Qi:2016}.
In particular, for $d=4$ the upper bound is visible in the random sample in Figure \ref{fig:gyration_radius}.

The lower bound of (\ref{eq:Kirchhoff_bounds}) is not saturating. It is for instance assumed by the complete graph $K_v$, in which all $v$ vertices are connected to each other (requiring $v = d + 1$) \cite{Palacios:2016}. Generally, for highly connected graphs the value of the Kirchhoff index is at the lower end of the distribution. But since the lower bound is not saturating, one might wonder whether the actual minimal values are of order $\cO(1)$ for large $v$ or maybe growing logarithmically. This possiblity can be ruled out by the fact that the average value of the gyration radius for large $v$ is finite, as will be shown further down in this section, and therefore the lower bound in the infinite limit must also be finite, that is of order $\cO(1)$.

We found that for large $v$ highly connected graphs at the lower bound have a gyration radius $R^2_\gyr(\gamma) = \alpha D\cdot \cO(1)$ and therefore are localized in a volume of size $\sim \sqrt{\alpha D}$. Graphs at the upper bound with long chains or large trees have gyration radius $R^2_\gyr(\gamma) = \alpha D\cdot \cO(v)$ and spread over a volume of size $\sim \sqrt{\alpha v D}$.

To estimate the average size of a vacuum bubble in the matrix model, let us see which behavior dominates according to the weights in the perturbative expansion in $\hat w$. As the average size of a vacuum bubble we define the expectation value for the gyration radius with respect to the free energy (\ref{eq:free_energy_w}),
\begin{equation*}
	R^2_\gyr := \langle R^2_\gyr(\gamma) \rangle_{\hat w} = 
	\alpha D \left\langle r^2_\gyr(\gamma) \right\rangle_{\hat w}\, .
\end{equation*}
Let us perform a qualitative estimation of this quantity.

The weight for a given graph $\gamma$ in $\hat w$ is determined by the refined generalized Catalan numbers in the graph polynomial $\cP_{\bfd,\gamma}(N^{-2})$ and the weight factor from the reduced Laplacian determinant $(\det \Delta')^{-D/2}$. This suggests that the large $v$ behavior of the gyration radius $R^2_\gyr$ depends on $N$ and the dimension $D$, giving rise to a phase diagram in the $(D,N)$ plane.
A similar phase diagram has been studied in the context of triangulated random surfaces in \cite{Boulatov:1986jd}.\footnote{In \cite{Boulatov:1986jd} a $\phi^3$ triangulation model was investigated, the dual graphs having trivalent vertices. Here we consider graphs with fixed vertex degree $d$, that is $d$-regular graphs. Also, a different phase space was considered in the reference, in particular without $N$-dependence. To match notation, note that $v|_{\mathrm{here}} = 2N|_{\mathrm{BKKM}}$. $\alpha|_{\mathrm{BKKM}}$ is a coupling constant to a 2d curvature squared term on the triangulated world sheet and is not related to $\alpha|_{\mathrm{here}}$.} 

Although the refined generalized Catalan numbers are not known explicitly, picking a test sample of regular graphs and plotting the gyration radius $r^2_\gyr$ over the vertex number, see Figure \ref{fig:gyration_radius}, shows that the distribution of graphs peaks around values of the gyration radius of order $\cO(1)$. That is, for growing vertex number highly connected graphs dominate the distribution. This is the governing behavior for vanishing or small dimension $D$ and for all genera having equal weight, that is $N=1$.

The value for large $v$ can be made precise using a quite remarkable result from graph theory, going back to 
Kesten and McKay \cite{Kesten:1959srw,McKay:1981eed}. It states that in the infinite $v$ limit the Eigenvalue distribution of the adjacency matrix for a randomly chosen $d$-regular connected graph is given by
\begin{eqnarray}
	\label{eq:Kesten-McKay}
	\rho_d(\mu) = \frac{d\sqrt{4(d-1)-\mu^2}}{2\pi(d^2-\mu^2)}\, .
\end{eqnarray} 
This distribution has support on the interval $[-2\sqrt{d-1},2\sqrt{d-1}]$. Note that for a $d$-regular graph the Eigenvalues $\{ \mu_i \}$ of the adjacency matrix $A$ are related to the ones of the Laplacian matrix by the shift $\lambda_i = d - \mu_i$.\footnote{The single vanishing Eigenvalue of the Laplacian matrix for connected graphs, that is $\mu_1 = d$ for $A$, is an exception and not taken into account in the distribution function (\ref{eq:Kesten-McKay}).}

To make contact with the gyration radius for large $v$ we use the Stieltjes transform of the Kesten-McKay distribution, which was explicitly derived in \cite{Dubbs:2015irmt},
$$
  S_d(z) = \int d\mu \, {\frac{\rho_d(\mu)}{z - \mu}} = \frac{2(d-1)}{(d-2)z + d\sqrt{z^2+4(1-d)}}.
$$
In fact, the gyration radius for the randomly chosen infinite graph becomes
$$
  r^2_\gyr(\gamma) = \frac{1}{v} \sum_{i=2}^v \frac{1}{d - \mu_i} 
  \longrightarrow \int d\mu\, \frac{\rho_d(\mu)}{d - \mu} = S_d(d) = \frac{d-1}{d(d-2)}\, .
$$ 
The random sample in Figure \ref{fig:gyration_radius} indeed verifies this limiting value, $S_4(4) = 3/8$. 
In summary, this result states that the gyration radius $r^2_\gyr$ for $D=0$ and $N=1$ is of order $O(1)$.

As $D$ grows the weight factor $(\det \Delta')^{-D/2}$ becomes increasingly relevant for the distribution of graphs over the gyration radius. Cycle graphs or tree graphs have few spanning trees and therefore a small value of the determinant, whereas highly connected graphs have a large number of spanning trees. 
This implies that with increasing dimension $D$ the weight factors of graphs with long chains, cycles or trees become increasingly relevant as compared to the weight factors for highly connected graphs, so that the gyration radius will be pulled to larger values, see Figure \ref{fig:gyration_radius}. 

The phase diagram in the $(D,N)$ plane can therefore have the following types of phases (cf. \cite{Boulatov:1986jd}): In a phase, where highly connected graphs dominate we expect that the gyration radius behaves as $R^2_\gyr \sim \alpha D$. In a branched polymer phase, where cycle graphs or trees dominate, the expectation is that $R^2_\gyr \sim a D$, where $a = \alpha \langle \hat v\rangle_{\hat w}$. Between these two limiting cases, phases can occur, where the gyration radius grows with a power law, $R^2_\gyr \sim \alpha D \langle \hat v^b \rangle_{\hat w}$ with $0 < b < 1$. Also, log-like growth can occur \cite{Boulatov:1986jd}.

In the present work we can only leave it with this qualitative observation, a quantitative or even analytic characterization of $R^2_\gyr$ for the transition from low to high $D$ and with varying $N$ is out of reach.

\subsection{Continuum limit}
\label{sec:ContinuumLimit}

In the previous section we studied the behavior of the gyration radius for large expectation values of the vertex number $\langle \hat v\rangle_{\hat w}$. In this section we review how to tune the coupling constant of the matrix model to reach this continuum limit.

We restrict our considerations to the quartic potential $V(M) = t_4 \trace (M^4)$. Note that in the quartic case we have $e = 2v$, and we can apply a field redefinition to set $t_4 = 1/\lambda$. For large vertex numbers the genus $h$ contribution to the free energy behaves as \cite{Ginsparg:1993is}
$$
  \cF_h \sim
  \sum_v v^{(\gamma_{\mathrm{str}}-2)\chi/2-1} (\lambda/\lambda_c)^v \sim 
	\left(1-\lambda/\lambda_c \right)^{(1-\gamma_{\mathrm{str}}/2)\chi}\, ,
$$
where $\lambda_c$ is a critical value for $\lambda$. For the critical limit $\lambda \rightarrow \lambda_c$ the string susceptibility $\gamma_{\mathrm{str}}$ is assumed to satisfy $\gamma_{\mathrm{str}}<2$. Generally it is hard to determine the explicit dependence of the string susceptibility on the dimension $D$. The result from Liouville theory is given by $\gamma_{\mathrm{str}} = 1/12\left(D-1-\sqrt{(D-1)(D-25)}\right)$ \cite{Knizhnik:1988ak,David:1988hj,Distler:1988jt}. This expression is real only for $D\leq 1$ and $D\geq 25$. Particularly, for $D\leq 1$ there are well-established results that relate the critical behavior of random matrix models to Liouville theory \cite{Kazakov:1988ch,Brezin:1990rb,Das:1990kaa,Kostov:1999xi}.

The appearance of the Euler character $\chi = 2-2h$ in $\cF_h$ suggests combining $N$ with $\lambda$ in the coupling constant
$$
  \kappa^{-1} := N (1-\lambda/\lambda_c)^{1-\gamma_{\mathrm{str}}/2}\, ,
$$
and taking a double scaling limit by keeping $\kappa$ fixed, while $N\rightarrow \infty$ and $\lambda \rightarrow \lambda_c$.  Using (\ref{eq:free_energy_w}) and the redefinition $\hat w_h = (1-\lambda/\lambda_c)^{(1-\gamma_{\mathrm{str}}/2)\chi} \bar w_h$ for each genus contribution, the free energy in the double scaling limit becomes 
$\hat\cF = V_\cM / (2\pi\alpha)^{D/2} \bar w(\kappa)$ with $\bar w(\kappa) = \sum_h \kappa^{2h-2} \bar w_h$.	

To see that the critical limit $\lambda \rightarrow \lambda_c$ corresponds to a continuum limit we consider the quantity $a = \alpha \left< \hat v \right>_{\hat w}$. The expectation value $\left< \hat v \right>_{\hat w}$ measures the average number of particle interactions involved in a vacuum bubble, that is, the area $a$ is a measure for the involved degrees of freedom.
For the assumed quartic potential and the assumption that $\gamma_{\mathrm{str}} < 2$, the area $a$ behaves as
$$
  a = \alpha \left< \hat v \right>_{\hat w} = 
    \alpha \frac{\lambda \partial_\lambda \hat w}{\hat w} =
	\alpha \left(
	\left(1-\frac{\gamma_{\mathrm{str}}}{2}\right)
	\left< 2 h - 2 \right>_{\hat w}
	\frac{\lambda/\lambda_c}{1-\lambda/\lambda_c}
	+ \frac{\lambda \partial_\lambda \bar w(\kappa)}{\bar w(\kappa)}
	\right)
	\, .
$$
In the critical limit the denominator of the first term on the right-hand side vanishes and the expectation value $\left< 2 h - 2 \right>_{\hat w}$ becomes infinite. So, keeping the expectation value for $a$ fixed, the critical limit sends $\left< \hat v \right>_{\hat w} \rightarrow \infty$ and $\alpha \rightarrow 0$, which corresponds to a continuum limit. Notice that as a consequence of the discussion in the previous section, the area $a$ is not necessarily the size of the vacuum bubble as embedded in the target space $\cM$. The average size of the vacuum bubble is given by the gyration radius $R^2_\gyr$.

For the special value $\gamma_{\mathrm{str}} = 2$ of the string susceptibility a separate consideration of the continuum limit is required. The genus $h$ contribution to the free energy has log-behavior
\begin{equation*}
  \cF_h \sim \sum_v \frac{1}{v}(\lambda/\lambda_c)^v
	\sim -\ln (1-\lambda/\lambda_c) \, ,
\end{equation*}
identical for all genera, so that a redefinition of the coupling constant in terms of $\kappa$ does not make sense. Rather, let us write $\hat w_h = -\ln (1-\lambda/\lambda_c)\, \bar w_h$ and $\bar w(N) = \sum_h N^{2-2h} \bar w_h$ and consider
$$
  a = \alpha \left< \hat v \right>_{\hat w} = 
    \alpha \frac{\lambda \partial_\lambda \hat w}{\hat w} =
	\alpha \left(
	\frac{1}{-\ln (1-\lambda/\lambda_c)}
	\frac{\lambda/\lambda_c}{1-\lambda/\lambda_c}
	+ \frac{\lambda \partial_\lambda \bar w(N)}{\bar w(N)}
	\right)
	\, .
$$
In the critical limit $\lambda \rightarrow \lambda_c$ the denominator of the first term on the right-hand side vanishes. So, we can again keep the value for $a$ fixed, and take the continuum limit sending $\left< \hat v \right>_{\hat w} \rightarrow \infty$ and $\alpha \rightarrow 0$.

\section{The matrix model in curved background}
\label{sec:MMCB}

Let us now consider the matrix model (\ref{eq:higherdim_matrix_model}) in a curved background with Riemannian metric $g$ and derive its free energy perturbatively. For what follows, it is important to notice that we do not impose any on-shell condition on the metric. Here and in the following we assume that the Riemannian manifold $\cM$ is simply connected and $\alpha$ is much smaller than the size of $\cM$, so that contributions from homotopically non-trivial loops do not occur and a WKB approximation applies. 

In the functional integral (\ref{eq:higherdim_matrix_model}) the Riemannian metric has the effect that the Laplacian differential operator and the integration measure must be covariant with respect to the metric $g$.
The action functional becomes
$$
		S(\alpha, \lambda, t_d, M(x), g_{\mu\nu}) = 
		\frac{N}{(2 \pi \alpha)^{D/2}}
		\int_\cM {d^Dx \sqrt{g} 
		\left( \frac{1}{2\lambda} \trace M(x) 
		e^{-\frac{\alpha}{2}\Delta_g} M(x) + V(M)
		\right)},
$$
where $g = \det{g_{\mu\nu}}$, and $\Delta_g = g^{\mu\nu}D_\mu D_\nu$ is the Laplacian in curved background.

The Green's function $G_g(\alpha,x,x')$ is defined in a covariant manner by
$$
	e^{-\frac{\alpha}{2}\Delta_g} G_g(\alpha,x,x') = 
	g^{-1/2}\, \delta^D(x-x') \ .
$$
Again this Green's function is, at the same time, the heat kernel to the heat equation in curved space and therefore satisfies
\begin{equation*}
	\begin{aligned}
		\partial_\alpha G_g(\alpha, x, x') &= \frac{1}{2} \Delta_g G_g(\alpha, x, x'), \\
		G_g(0,x,x') &= g^{-1/2}\, \delta^D(x-x').
	\end{aligned}
\end{equation*}

The heat kernel in curved background can be deduced from the Schwinger-DeWitt method following their seminal works \cite{Schwinger:1951nm, DeWitt:1975ys}. The ansatz for the Green's function is
\begin{equation}
  G_g(\alpha, x, x') = \frac{1}{(2 \pi \alpha)^\frac{D}{2}}
	                   \sqrt{\bfD(x,x')}
	                   e^{-\frac{\sigma(x,x')}{\alpha}} 
										 \, \Omega(\alpha,x,x')\ , 
  \label{eq:curved_Greens_function}
\end{equation}
where $\sigma(x,x')$ is one half of the geodesic distance squared between $x$ and $x'$, and $\bfD(x,x')$ is the Van Vleck-Morette determinant
\begin{equation}
  \label{eq:VVM_determinant}
  \bfD(x,x') = \frac{\det{(-\partial_\mu \partial'_\nu \sigma(x,x'))}}{\sqrt{\det g}\, \sqrt{\det g'}}\ .
\end{equation}
Expanding the function $\Omega(\alpha,x,x')$ as perturbative series in $\alpha$,
$$
	\Omega(\alpha,x,x') = 1 +
	\sum_{n=1}^\infty a_n(x,x') \left(\frac{\alpha}{2}\right)^n\ ,
$$
the coefficient functions $a_n(x,x')$ can be recursively determined by the heat equation.

Since we assume $\alpha$ to be small and because of the Gaussian nature of the Green's function it suffices to expand the coefficient functions around the coincidence limit. The leading term in the first coefficient is known to be \cite{DeWitt:1975ys, Gilkey:1975iq, Vassilevich:2003xt, Barvinsky:2019spa, Barvinsky:2021ijq}
$$
  a_1(x,x') = \frac{1}{6} R(x') + \cO(x-x')\ ,
$$
where $R$ is the scalar curvature of the metric $g$. The higher order coefficients, $a_n(x,x')$ for $n \geq 2$, carry curvature squared and higher terms in the coincidence limit. The $\cO(x-x')$-terms, corresponding to derivatives of the curvature, turn out to contribute to higher orders in $\alpha$ after integration.

Following the same steps as in the previous section to arrive at the perturbative expansion for the free energy we find the approximate result
\begin{eqnarray}
	\label{eq:free_energy_curved_gaussian}
	\hat \cF &=& \sum_{h=0}^\infty N^{2-2h} 
	\sum_{\mathbf{d} \in Y_{\textit{ev}}}(-1)^{v+1}\, 
	\frac{\lambda^e\, t_{\mathbf{d}}}{S(\mathbf{d})} \times \\
	&\times& \hspace{-10pt}\sum_{\Gamma\in \hat G_{h}(\mathbf{d})}
	\int{\frac{d^D_g X}{\left(2\pi \alpha\right)^{Dv/2}} 
	  \prod_{e_{ij}} \left(\sqrt{\bfD(x_i, x_j)}
	  e^{- \frac{\sigma(x_i,x_j)}{\alpha}} \left(1 + \frac{\alpha}{12} R(x_j)\right) 
	  \right)}\, 
\end{eqnarray}
Using the expansion (\ref{eq:Morette_expansion}) of the van Vleck-Morette determinant, one sees that this is a discrete string perturbation expansion (when taking the metric to be on-shell).

\begin{figure*}[!t]
  \centering
	\begin{tabular}{cc}
		\includegraphics[width=0.46\textwidth]{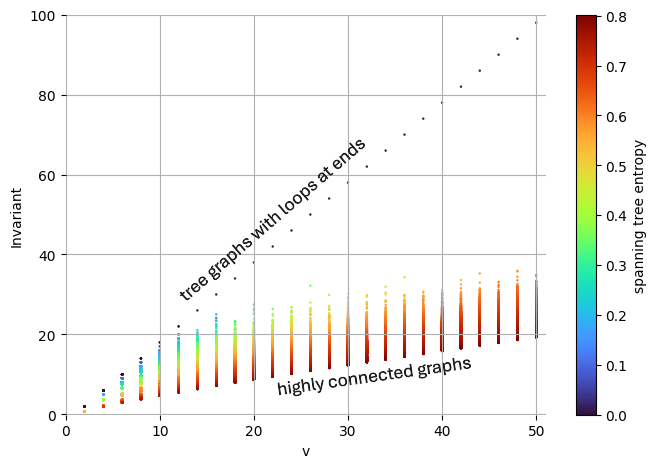} &
		\includegraphics[width=0.46\textwidth]{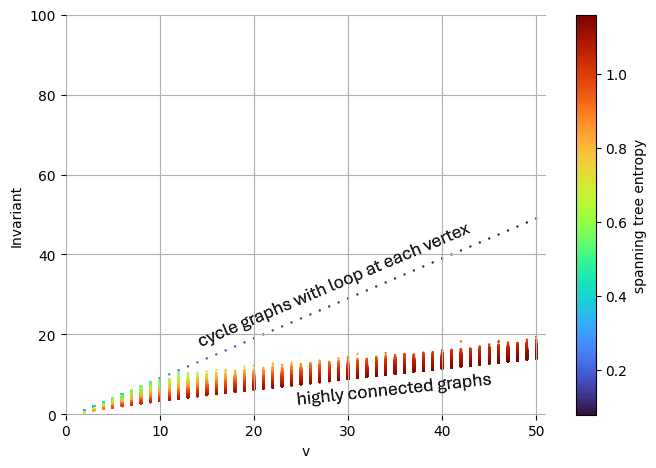}  \\
		\small{(a) $\cI$ for $3$-regular graphs} & \small{(b) $\cI$ for $4$-regular graphs}
	\end{tabular}
	\caption{A random sample of regular graphs shows that the gravitational graph invariant $\cI$ lies within linear bounds for large $v$.}
	\label{fig:invariant}
\end{figure*}

The embedding coordinates can be integrated out because of the Gaussian nature of the integrands, when we assume the metric $g$ to vary slowly as compared to the gyration radius. All contributing heat kernels are expanded around a reference coordinate $x=x_v$, using the coordinates $x'_i = x_i - x$ for $i=1,\ldots,v-1$. The computation is outlined in Appendix \ref{apdx:heat_kernel_curved} and leads to the free energy
\begin{eqnarray*}
	\hat \cF(\alpha, \lambda, t_d, N) = 
	\frac{1}{(2\pi\alpha)^{D/2}} 
	\sum_{h=0}^\infty N^{2-2h}
	\hspace{-25pt}&&
	\sum_{\mathbf{d} \in Y_{\textit{ev}}}(-1)^{v+1}
  \frac{\lambda^e\, t_{\bfd}}{S(\bfd)} 
	\sum_{\gamma\in Sk_{\mathbf{d}}}
	\frac{C_{h}(\bfd, \gamma)}{\det\Delta'(\gamma)^{D/2}} \times\\
	 &\times& \int d^D_g x \left( 1 + \frac{\alpha}{12}(e + v - 1 - \cI) R(x)\right).
\end{eqnarray*}
As a result of this derivation the quantity $\cI = \cI(\gamma)$ must be a graph invariant. It is given by
\begin{equation}
  \label{eq:graph_invariant}
	\cI(\gamma) = 2 \sum_i \Delta'^{-1}_{ii} + 
	\sum_{i,j} \left(
		\Delta'^{-1}_{ii}\, \Delta'_{ij}\,  \Delta'^{-1}_{jj}
		-\Delta'^{-1}_{ij} \Delta'_{ij}\,  \Delta'^{-1}_{ij}
	\right)\, .
\end{equation}
Let us henceforth refer to it as gravitational graph invariant.
Mathematically, it remains a conjecture that $\cI$ is a graph invariant, but an explicit calculation on a random sample of Laplacian matrices (cf. Figure \ref{fig:invariant}) verifies that $\cI$ is independent of the choice of the reduction of the Laplacian matrix $\Delta$ to $\Delta'$, which supports the conjecture.

Recalling the combinatorial expression (\ref{eq:FreeEnergyInvariantPolynomials}) for $\hat w = \hat w(\lambda, t_d, N)$, the free energy can be concisely written as Einstein-Hilbert action with cosmological constant term,
\begin{equation}
	\label{eq:Einstein}
	\hat\cF(\alpha, \lambda, t_d, N) = 
	\frac{1}{(2\pi\alpha)^{D/2}}
	\int d^D x \sqrt{g} \, \hat w \left( 1 + 
	\frac{\alpha}{12}\, 
	\langle\, \hat\cV\, \rangle_{\hat w}\, R(x)\right)
	\, ,
\end{equation}
where $\langle \hat\cV \rangle_{\hat w}$ is the expectation value
\begin{equation}
	\label{eq:total_graph_invariant}
    \langle \hat\cV \rangle_{\hat w} = \langle \hat e + \hat v - 1 - \hat \cI \rangle_{\hat w}\ .
\end{equation}

From (\ref{eq:Einstein}) the gravitational constant $\kappa = 8\pi G$ can be read off as
\begin{equation*}
	\kappa^{-1}
	= \frac{\hat w}{(2\pi\alpha)^{D/2}}
	\frac{\alpha}{6} \, \langle \hat\cV \rangle_{\hat w}.
\end{equation*}
The cosmological constant is given by
\begin{equation*}
	\Lambda^{-1} = \frac{\alpha}{6} \, \langle \hat\cV \rangle_{\hat w}
	\, ,
\end{equation*}
or in terms of the vacuum energy density by
\begin{equation*}
	\rho_{\mathrm{vac}} = \Lambda / \kappa = \frac{\hat w}{(2\pi\alpha)^{D/2}}
	\, .
\end{equation*}
Note that the cosmological and the gravitational constant are expressed through $\hat w$ and the expectation value $\langle \hat\cV \rangle_{\hat w}$. Both are solely determined by graph combinatorics, and notably contain the full string perturbation expansion to all orders in genus $h$.

To consider the behavior of $\alpha \langle \hat\cV \rangle_{\hat w}$ in the continuum limit we need to study the gravitational graph invariant $\cI$ more closely. Figure \ref{fig:invariant} shows how $\cI$ is distributed over the vertex number $v$ for a random sample of regular graphs of degree $3$ and $4$. This motivates the conjecture that for regular graphs of degree $d$ the invariant is linearly bounded as follows:
\begin{eqnarray*}
	\frac{v}{d} ~\leq~ \cI(\gamma) ~\leq~
	\left\{
		\begin{array}{cl}
			v  & \textrm{for}~d~\textrm{even}, \\ 
			2v & \textrm{for}~d~\textrm{odd}.
		\end{array}
	\right.
\end{eqnarray*}
These bounds in particular imply that the invariant grows linearly for large $v$, that is
$$
  \langle \hat\cI \rangle_{\hat w} \rightarrow C_d \langle \hat v \rangle_{\hat w}\, .
$$
The constant $C_d$ must respect the above bounds and depends on the degree $d$ as well as the dimension $D$ and $N$.

A matrix model with monomial potential $V(M)$ of degree $d$ therefore gives rise to a cosmological constant in the continuum limit as
$$
  \Lambda^{-1} 
  \rightarrow 
  \frac{\left(d + 2 - 2C_d\right)}{12} \, a \, .
$$
This result relates the cosmological constant to the area $a = \alpha \langle \hat v\rangle_{\hat w}$ measuring the degrees of freedom of the vacuum bubble. The coefficient $d+2-2C_d$ is always positive given the above bounds.

A large cosmological constant implies small values for $a$ and consequently for $R^2_\gyr \lesssim a D$. The assumption for deriving the Einstein-Hilbert action, that $R^2_\gyr$ is smaller than typical changes of the metric $g$, can therefore be satisfied for sufficiently large $\Lambda$.

Recall however that we argued in Section \ref{sec:Gyration} that the large $v$ behavior of the gyration radius depends on the phase in the $(D,N)$ plane. Only in the branched polymer phase, where $R^2_\gyr \sim a D$, a sufficiently small gyration radius also requires $a$ to be small and $\Lambda$ to be large. In all other phases, taking the continuum limit with $\langle \hat v\rangle_{\hat w} \rightarrow \infty$ and $\alpha \rightarrow 0$, keeping $a$ fixed, the gyration radius $R^2_\gyr$ tends to zero. As a consequence $a$ can be set to an arbitrary finite value.

In fact, the observed cosmological constant $\Lambda_{\mathrm{obs}}$ (ignoring for a moment the dimension of $\cM$ and the fact that we consider a Riemannian manifold) is an extremely tiny quantity, which can be related to a characteristic length scale $\ell_{\mathrm{obs}} \sim \Lambda^{-1/2}_{\mathrm{obs}}$, which curiously turns out to be of the order of the observable universe. Outside the branched polymer phase, such a large but finite length scale is consistent with the approximation that we required for deriving the Einstein-Hilbert action.

\section{The matrix model for ribbon graphs with boundary}
\label{sec:VMM}

Let us return to the Euclidean background with $\cM=\bbR^D$ and extend the matrix model in such a way that it generates ribbon graphs with boundary components. To this end we couple the matrix $M$ to an $n$-tuple of $N$-dimensional vector fields, subsumed in the complex-valued $n \times N$-matrix $B$. By abuse of language, we will refer to $B$ as vector field.

While the matrix $M$ is defined on $\cM = \bbR^D$, the vector field $B$ and its coupling to $M$ may be restricted to a $d$-dimensional subplane $\cS \subset \cM$ with $d \leq D$. Without loss of generality the plane is rotated and shifted, so that $\cS = \{x_{d+1}=\ldots=x_D=0\}$. For later clarity the coordinates along $\cS$ are denoted by $(y_1,\ldots,y_d)$, whereas the coordinates normal to $\cS$ are denoted by $(z_{d+1},\ldots,z_D)$.
The action for the vector-valued field $B$, including the coupling to the matrix $M$, is defined by
$$
  S_B = 
	\frac{N}{(2\pi\alpha)^{d/2}} \int d^dy \, 
	\frac{1}{2\lambda} \bar B(y) e^{-\frac{\alpha}{2} \Delta} B(y) +
	s \,  \bar B(y) M(y) B(y) \, ,
$$
with real coupling constant $s$, and $\bar B = B^\dagger$. The matrix model partition function is
\begin{equation}
  \label{eq:VMM}
	Z = \int \cD  M(x)\, \cD  B(y)\, e^{-S_E - S_B}\, .
\end{equation}
The propagator for the vector field is the same as for the matrix field with the only difference that it is restricted to the subplane $\cS$:
$$
  G_B(\alpha, y,y') = \frac{1}{(2\pi\alpha)^{d/2}} \exp\left(-\frac{(y-y')^2}{2\alpha}\right)\, .
$$

Working out the perturbative expansion, the contributions to the free energy that do not originate from vector fields remain unchanged and are given by (\ref{eq:FreeEnergyInvariantPolynomials}). The contributions that do involve the vector-valued field lead to a perturbative expansion over connected ribbon graphs with $b$ boundary components.

In the following the number of boundary vertices and edges for each boundary component are denoted by $v_a$ and $e_a$ for $a=1,\ldots,b$, and we use the abbreviation $\bfv_B = (v_1,\ldots,v_b)$. Notice that for each boundary component, we have an equal number of boundary vertices and edges, that is $v_a=e_a$ for $a=1,\ldots,b$. The total number of boundary vertices and edges are $v_B = \sum_a v_a$ and $e_B = \sum_a e_a$, respectively. The number of inner vertices and edges are $v_I$ and $e_I$, giving the total numbers $v = v_I + v_B$ and $e = e_M + e_B$. The Euler formula becomes
$$
	2-2h-b = v-e+f = v_I-e_I+f.
$$
The free energy for graphs containing at least one boundary component is
\begin{align}
	\label{eq:openclosedeffaction}
  \hat \cF_B(\alpha, \lambda, t_d, s, N) =& 
	\sum_{h=0}^\infty  
	\sum_{(\bfd,\bfv_B) \in (Y, Y)_{\textit{ev}}} \hspace{-10pt}
	N^{2-2h-b} n^b\;
	(-1)^{v+1}\frac{\lambda^{e}\, t_{\bfd}\, s^{v_B}}
	{S(\mathbf{d}) S(\bfv_B)} \times \\ \nonumber
	\times &\sum_{\Gamma\in \hat G_{h,b}(\bfd,\bfv_B)} \hspace{-10pt}
	\int \frac{d^{dv} y}{\left(2\pi \alpha\right)^{dv/2}}{
	e^{-\frac{1}{2\alpha} (\sum_{e_{ij}} (y_i-y_j)^2+
	\sum_{e_{ip}} (y_i-y_p)^2+\sum_{e_{pq}} (y_p-y_q)^2)}
	}\\ \nonumber
	\times&\int \frac{d^{(D-d)v_I} z}{\left(2\pi \alpha\right)^{(D-d)v_I/2}}{
	e^{-\frac{1}{2\alpha} (\sum_{e_{ij}} (z_i-z_j)^2+\sum_{e_{ip}} z_i^2)}
	}\, .
\end{align}
$Y$ is the set of Young diagrams, and $(Y,Y)_{\textit{ev}}$ is the set of Young diagram tuples of even total degree. The first Young diagram $\bfd$ denotes the degrees of inner vertices (from the interaction term in $S_E$), and the second Young diagram $\bfv_B$ denotes the number of vertices of each boundary component (from the interaction in $S_B$). The number of boundary components $b$ is implicitly given by the number of rows of the Young diagram $\bfv_B$. 
$e_{ij}$ are edges between two inner vertices, $e_{ip}$ edges between inner and boundary vertices, and $e_{pq}$ edges between two boundary vertices.
$\hat G_{h,b}(\bfd, \bfv_B)$ is the set of decorated marked ribbon graphs for genus $h$ and $b$ boundary components. 

For the ribbon graph $\Gamma$ with $b$ boundary components, let $\gamma$ be its skeleton, forgetting cyclic ordering, marking and decoration: The map from the set of decorated marked ribbon graphs $\hat G_{h,b}(\bfd,\bfv_B)$ to unmarked skeletons $Sk(\bfd,\bfv_B)$,
\begin{eqnarray*}
  \operatorname{sh}_{h,b}: G_{h,b}(\bfd,\bfv_B) &\rightarrow& Sk_b(\bfd,\bfv_B) , \\
	\Gamma &\mapsto & \gamma ,
\end{eqnarray*}
is defined by forgetting the cyclic order of half-edges at each vertex, but keeping the information whether a vertex is an inner or a boundary vertex.

To perform the integration, we rewrite the integrand using the Laplacian matrix $\Delta(\gamma)$ for the skeleton $\gamma$. The restriction of the Laplacian matrix to inner vertices is denoted by $\Delta_I(\gamma)$ and has full rank. As before, $\Delta'(\gamma)$ is the reduced Laplacian. Performing a coordinate transformation $(y,Y',Z)$ that makes the translational invariance explicit along $y$, we rewrite the quadratic expressions in the exponents of the Gaussian distribution as $Y' \Delta' Y'$ and $Z \Delta_I Z$. This allows us to integrate over the embedding coordinates, thus arriving at
\begin{align*}
  \hat \cF_B(\alpha, \lambda, t_d, s, N) =&
	\frac{V_\cS}{(2\pi\alpha)^{d/2}}
	\sum_{h=0}^\infty  
	\sum_{(\bfd,\bfv_B) \in (Y, Y)_{\textit{ev}}} \hspace{-10pt}
	N^{2-2h-b} n^b\;
	(-1)^{v+1}\frac{\lambda^{e}\, t_{\mathbf{d}}\, s^{v_B}}
	{S(\mathbf{d}) S(\bfv_B)}\\
	\times&\sum_{\Gamma\in \hat G_{h,b}(\mathbf{d},\bfv_B)}
	\frac{1}
	{{\det\Delta'(\gamma)}^{d/2} \det\Delta_I(\gamma)^{(D-d)/2}}\, ,
\end{align*}
where $V_\cS = \int d^d y$ is the infinite volume of $\cS$.

For a given skeleton $\gamma \in Sk_b(\bfd,\bfv_B)$, let us denote the number of decorated marked ribbon graphs by
$$
  C_{h,b}(\bfd, \bfv_B, \gamma) := 
	\sum_{\Gamma \in \operatorname{sh}_{h,b}^{-1}(\gamma)}
	C_{h,b}(\bfd, \bfv_B, \Gamma),
$$
and introduce the invariant polynomial for the skeleton $\gamma$ by
$$
  \cP_{\bfd,\bfv_B,\gamma}(N^{-1}) := \sum_{h} N^{-2h-b} \, C_{h,b}(\bfd,\bfv_B,\gamma).
$$

Putting all parts of the free energy together we obtain
$$
  \hat \cF(\alpha, \lambda, t_d, s, N) = 
	\frac{V_\cM}{(2\pi\alpha)^{D/2}} \hat w(\lambda, t_d, N) + 
	\frac{V_\cS}{(2\pi\alpha)^{d/2}} \hat w_B(\lambda, t_d, s, N) \, . 
$$
$\hat w(\lambda, t_d, N)$ is given by (\ref{eq:FreeEnergyInvariantPolynomials}), and the perturbation expansion for ribbon graphs with boundary is subsumed in
\begin{equation*}
	\hat w_B(\lambda, t_d, s, N) =
	\hspace{-15pt}\sum_{(\bfd,\bfv_B) \in (Y, Y)_{\textit{ev}}} 
	\hspace{-10pt}
	(-1)^{v+1}\frac{\lambda^{e}\, t_{\mathbf{d}}\, s^{v_B}}
	{S(\mathbf{d}) S(\bfv_B)}
	\sum_{\gamma\in Sk(\mathbf{d},\bfv_B)}
	\frac{n^b N^2\, \cP_{\bfd,\bfv_B,\gamma}(N^{-1})}
	{{\det\Delta'(\gamma)}^{d/2} \det\Delta_I(\gamma)^{(D-d)/2}}
	\, .
\end{equation*}

For the vector matrix model that gives rise to ribbon graphs with boundary, we therefore find that the cosmological constant term with $\hat w$ is still distributed over $\cM$, whereas the term with $\hat w_B$  localizes on the subplane $\cS$. In the continuum interpretation, the subplane corresponds to a D-brane.

\section{The matrix model coupled to a background gauge field}
\label{sec:MMGF}

Our next step is to consider the matrix model (\ref{eq:VMM}) in the presence of a non-trivial gauge field background on $\cS$. For simplicity, we stay on a Euclidean background $\cM = \bbR^D$.

The $N$-dimensional complex vector field $B$ on the subplane $\cS = \bbR^d$ is assumed to be minimally coupled to a non-Abelian background gauge field $A$ for the gauge group $U(n)$, 
$$
  D^A_\alpha B(y) = \left(\partial_\alpha - i A_\alpha\right) B(y)\, .
$$
The Laplacian operator acting on $B$ is accordingly extended by the gauge field through
$$
  \Delta_A B(y) = D^A_\alpha D^A_\alpha B(y)\, .
$$

The action for $B$, including the coupling to the matrix $M$, is
$$
  S_B = 
	\frac{N}{(2\pi\alpha)^{d/2}} \int d^dy \, 
	\frac{1}{2\lambda} \bar B(y) e^{-\frac{\alpha}{2} \Delta_A} B(y) +
	s \,  \bar B(y) M(y) B(y) \, .
$$

The Greens function for the vector-valued field $B$ is determined by the solution to the heat equation for the Laplacian $\Delta_A$. It can be determined by the Schwinger-DeWitt heat kernel method \cite{Gilkey:1975iq, Vassilevich:2003xt, Barvinsky:2019spa, Barvinsky:2021ijq}. 
As an ansatz for the solution to this heat equation, the Gaussian distribution is dressed with a correction factor that can be expanded in $\alpha$:
$$
  G_A(y,y') = \frac{1}{(2\pi\alpha)^{d/2}} e^{-\frac{(y-y')^2}{2\alpha}} \Omega(\alpha,y,y')\, ,
$$
with
$$
  \Omega(\alpha, y,y') = \sum_{n=0}^{\infty} \left(\frac{\alpha}{2}\right)^n a_n(y,y')\, .
$$
The coefficients in the expansion come with various names, like HaMiDeW (Hadamard-Minakshisundaram-DeWitt) or Seeley-Gilkey heat kernel coefficients. In \cite{Gilkey:1975iq, Vassilevich:2003xt} these coefficients are computed in the coincidence limit $y=y'$. Barvinsky and Wachowski developed a method to determine the first HaMiDeW coefficients without coincidence limit \cite{Barvinsky:2021ijq}. 

For now, considering flat space and only the coincident limit except for the parallel transport by the Wilson line $W(y,y') := \exp{\left(i\int_{y'}^{y} A \right)}$, the Greens function is
\begin{equation*}
  G_A(y,y') = \frac{1}{(2\pi\alpha)^{d/2}} e^{-\frac{(y-y')^2}{2\alpha}} 
  W(y,y') \left(1 + \frac{\alpha^2}{48} F_{\mu\nu}F^{\mu\nu}(y')\right) \, .
\end{equation*}

Inserting this Greens function in the computation of the free energy gives, for every boundary component $\gamma_a$ (for $a=1,\ldots,b$) of
the graph $\gamma$, an additional factor under the integral in (\ref{eq:openclosedeffaction})
$$
  \prod_{a=1}^b \left(\trace W_{\gamma_a} + 
	\frac{\alpha^2}{48} \sum_{p\in V(\gamma_a)} 
	\trace \left(W_{\gamma_a}(y_{p}, y_{p}) F_{\mu\nu}F^{\mu\nu}(y_{p})  
	\right) \right) \, .
$$
The Wilson loop $\trace W_{\gamma_a}$ runs along the edges of the boundary component $\gamma_a$. 
The second term sums over the set of all vertices of the boundary component $\gamma_a$, and its Wilson line $W_{\gamma_a}(y_p,y_p)$ starts and ends at the point $y_p$ going once around the boundary component $\gamma_a$.

For a slowly varying field (as compared to the gyration radius) the Gaussian integrals in the free energy localize around the reference point $y$, and with $F_{\mu\nu} = F_{\mu\nu}(y)$ the new factor simplifies in leading order in $\alpha$ to be
$$
  n^b \left(1 + \frac{\alpha^2}{48} \frac{v_B}{n} \trace \left(F_{\mu\nu}F^{\mu\nu}\right)\right) \, . 
$$ 
Recall that $e_B = v_B$ is the total number of boundary edges or vertices.

The free energy for the gauge field background becomes
\begin{equation}
	\label{eq:free_energy_gauge_field}
  \cF(\alpha, \lambda, t_d, s, N) = 
	\frac{V_\cM}{(2\pi\alpha)^{D/2}} \hat w + 
	\frac{1}{(2\pi\alpha)^{d/2}} \int_\cS d^d y\, \hat w_B\left(
	1 + \frac{\alpha^2}{48 n} \left< \hat v_B \right>_{\hat w_B}
	\trace \left(F_{\mu\nu}F^{\mu\nu}\right)
	\right). 
\end{equation}
Both, the cosmological constant term, localized on the subplane $\cS$, and the Yang-Mills action come with coefficients determined by $\hat w_B$, which subsumes the combinatorics of the full discrete string perturbation expansion for ribbon graphs with boundary. Notice that the coupling of the gauge kinetic term is related to the expectation value of the boundary length,
$$
  \left< \hat v_B \right>_{\hat w_B} = 
	\frac{s \partial_s \hat w_B}{\hat w_B}\, .
$$
The gauge coupling is therefore determined by the average number of boundary interaction in the vacuum bubble.

\section{Covariant generalization}
\label{sec:covariant_generalization}

We finally place the matrix model (\ref{eq:VMM}) on a Riemannian manifold $\cM$ with metric $g$ ($\dim \cM = D$) and a submanifold $\iota: \cS \hookrightarrow \cM$ ($\dim \cS = d$) with induced metric $\bar g = \iota^*g$ and gauge field background. The Laplacian operator acting on the field $B$ is covariant with respect to the induced metric $\bar g$ and the gauge field. We assume that $\cM$ and $\cS$ are well-behaved enough so that the WKB approximation can be applied.

For a given ribbon graph $\Gamma$ with boundary, the weight factor in the perturbation expansion of the free energy is, up to $\alpha$-dependent prefactors, given by
$$
  \prod_i \int_{\cM} d^D x_i \sqrt{g(x_i)} 
  \prod_p \int_{\cS} d^d y_p \sqrt{\bar g(y_p)}
  \prod_{e_{ij}}G_g(\alpha, x_i, x_j) 
  \prod_{e_{iq}}G_g(\alpha, x_i, y_q) 
  \prod_{e_{pq}}G_{A,g}(\alpha, y_p, y_q) \, ,
$$
where $G_g$ is the heat kernel for the matrix $M$, $G_{A,\bar g}$ the one for $B$. $x_i$ are local coordinates on $\cM$, and $y_p$ local coordinates on $\cS$. 

According to \cite{Gilkey:1975iq, Vassilevich:2003xt, Barvinsky:2019spa, Barvinsky:2021ijq} the heat kernels for small $\alpha$ are given by
\begin{eqnarray*}
  G_g(\alpha, x_i, x_j) &=& 
	\frac{\sqrt{\bfD(x_i,x_j)}}{(2\pi\alpha)^{D/2}}
	e^{-\frac{1}{\alpha}\sigma(x_i, x_j)}
	\left(
		1 + \frac{\alpha}{12} R(x_j) 
	\right)\, , \\
  G_g(\alpha, x_i, y_q) &=& 
	\frac{\sqrt{\bfD(x_i,y_q)}}{(2\pi\alpha)^{D/2}}
	e^{-\frac{1}{\alpha}\sigma(x_i, y_q)}
	\left(
		1 + \frac{\alpha}{12} R(y_q) 
	\right)\, , \\
	G_{A,\bar g}(\alpha, y_p, y_q) &=& 
	\frac{\sqrt{\bar\bfD(y_p,y_q)}}{(2\pi\alpha)^{d/2}}
	e^{-\frac{1}{\alpha}\bar\sigma(y_p, y_q)}
	W(y_p,y_q)
	\left(
		1 + \frac{\alpha}{12} \bar R(y_q) + 
		\frac{\alpha^2}{48} F_{\alpha\beta}F^{\alpha\beta}(y_q) 
	\right)\, ,
\end{eqnarray*}
where $R$ is the curvature scalar of $g$ on $\cM$, and $\bar R$ is the curvature scalar of the induced metric $\bar g$ on $\cS$. $\bar \sigma$ is half the geodesic distance square and $\bar\bfD$ the Van Vleck-Morette determinant with respect to the induced metric $\bar g$. 

Because of the Gaussian behavior of the heat kernel, the integrals localize in a tubular neighborhood of $\cS$ in $\cM$, the size being controlled by $\alpha$. The tubular neighborhood of $\cS$ is diffeomorphic to the normal bundle $\cN\cS$, and it is possible to choose Fermi coordinates, $y$ along $\cS$ and $z$ in the fiber, in such a way that along $\cS$ the metric splits as $g|_{\cS} = \bar g \oplus g^\perp$, where $\bar g_{\alpha\beta}$ is the induced metric on $\cS$ and $g^\perp_{\alpha'\beta'}$ the metric on the fiber of the normal bundle $\cN\cS$. We use indices $\alpha, \beta, \ldots \in \{1,\ldots,d\}$ for coordinates along $\cS$, and $\alpha', \beta', \ldots \in \{d+1,\ldots,D\}$ in the normal direction.

In \cite{Graham:2024gnc} Graham and Kuo constructed normal coordinates adapted to the submanifold $\cS$ such that in the tubular neighborhood $\cN\cS$ the metric $g$ is expanded around $g|_{\cS} = \bar g \oplus g^\perp$, and the expansion coefficients of the metric are given by polynomials of the curvature tensor $R$, restricted to $\cS$, and the second fundamental form $B$ as well as their derivatives. Recall that the second fundamental form is defined by $B(X,Y) = D_X Y - \bar D_X Y$ and measures the rotation of a vector $Y$ in the tangent bundle $T\cS$ into the normal bundle $\cN\cS$ when parallel transported along the tangent vector $X$ on $\cS$.

In the adapted normal coordinates the curvature tensor $R$ decomposes according to the decomposition of the tangent bundle $T\cM|_\cS = T\cS \oplus \cN\cS$. The components $R_{\alpha\beta\gamma\delta}$, $R_{\alpha\alpha'\beta\beta'}$, $R_{\alpha'\beta'\gamma'\delta'}$ are important in the following, other components drop out because of symmetry reasons. 

Along the submanifold, the Gauss equation relates the curvature tensor $R$ for $g$ to both, the induced curvature tensor $\bar R$ and the second fundamental form $B$,
\begin{equation}
  \label{eq:Gauss_equation}
  R_{\alpha\gamma\beta\delta} =
	\bar R_{\alpha\gamma\beta\delta} +
	\bar g_{\alpha'\beta'} 
	(B^{\alpha'}_{\alpha\delta} B^{\beta'}_{\beta\gamma} -
	B^{\alpha'}_{\alpha\beta} B^{\beta'}_{\gamma\delta}
	)\, .
\end{equation}
As a consequence, the Ricci tensor decomposes accordingly,
\begin{eqnarray*}
  R_{\alpha\beta} &=& \bar R_{\alpha\beta} +
	(B_{\alpha'\alpha}{}^{\gamma} B^{\alpha'}_{\gamma\beta} - 
	B^{\alpha'}_{\alpha\beta} B_{\alpha'}) + 
	R^{\parallel\perp}_{\alpha\beta}\, ,\\
  R_{\alpha'\beta'} &=& R^{\parallel\perp}_{\alpha'\beta'} + 
	R^{\perp\perp}_{\alpha'\beta'}\, ,
\end{eqnarray*}
where 
\begin{equation}
	\label{eq:extrinsic_curv_defs}
  B_{\alpha'} = B_{\alpha'\alpha}^{\alpha}, \quad
	R^{\parallel\perp}_{\alpha\beta} = 	R_{\alpha\alpha'\beta}{}^{\alpha'}, \quad
  R^{\parallel\perp}_{\alpha'\beta'} = 
	R^{\alpha}{}_{\alpha'\alpha\beta'},\quad
  R^{\perp\perp}_{\alpha'\beta'} = 
	R_{\alpha'\gamma'\beta'}{}^{\gamma'}.
\end{equation}
And the relation for the curvature scalar is
$$
  R = \bar R + 
	(B_{\alpha'\alpha\beta} B^{\alpha'\alpha\beta} - B^{\alpha'} B_{\alpha'}) +
	2 R^{\parallel\perp} + R^{\perp\perp}\, .
$$

With these preparations the heat kernels can be expanded around a reference point on $\cS$, which is picked as one of the points $y_p$, say $y_{v_B}=y$. 
In Appendix \ref{apdx:heat_kernel_submfld} the expansion is outlined taking into account the coincidence limits of $\sigma(x,x')$ and the Van Vleck-Morette determinant \cite{Barvinsky:2021ijq}, as well as the expansion of the metric components in the adapted normal coordinates of \cite{Graham:2024gnc}.
Using the result (\ref{eq:int_over_heat_kernels}) and integrating out the embedding coordinates the final result for the free energy is in leading order of $\alpha$:
\begin{eqnarray}
  \label{eq:covariant_eff_action}
  \hat\cF &=& 
	\frac{1} {(2\pi\alpha)^{D/2}} \int_\cM d^D x \sqrt{g}\, 
	\hat w \left(1 + \frac{\alpha}{12} 
	\,\langle\, \hat\cV\, \rangle_{\hat w}\, R\right) +
	\\
	&+& \frac{1}{(2\pi\alpha)^{d/2}} \int_\cS d^d y \sqrt{\bar g}\,
	\hat w_B \Bigl(
	1 + \frac{\alpha}{12}
	\langle\, \hat\cV\, \rangle_{\hat w_B} \bar R +
	\frac{\alpha}{12}
	\langle\, \hat\cV_I\, \rangle_{\hat w_B} R^{\perp\perp} +
	\frac{\alpha^2}{48 n} \langle\,  \hat v_B \rangle_{\hat w_B}	\trace {F^2}
	\Bigr) + \nonumber \\
	&+& \frac{1}{(2\pi\alpha)^{d/2}} \int_\cS d^d y \sqrt{\bar g}\,
	\hat w_B \Bigl(
	    a_1\, B_{\alpha'\alpha\beta} B^{\alpha'\alpha\beta} +
		a_2\, B^{\alpha'} B_{\alpha'} +
		a_3\, R^{\parallel\perp}
	\Bigr)
	\, . \nonumber
\end{eqnarray}
The expectation values in the second line are taken with respect to $\hat w_B$ and are
\begin{eqnarray*}
	\langle\, \hat\cV\, \rangle_{\hat w_B} &=& 
	\bigl<\hat e + \hat v - 1 - \hat \cI\bigr>_{\hat w_B}\, , 
	\nonumber\\
	\langle\, \hat\cV_I\, \rangle_{\hat w_B} &=& 
	\bigl<\hat e_I + \hat v_I - \hat \cI_I\bigr>_{\hat w_B}\, .
\end{eqnarray*}
The graph invariant $\cI_I = \cI_I(\gamma)$ is given by
\begin{equation*}
	\cI_I(\gamma) = 2 \sum_i \Delta^{-1}_{I\,ii} + 
	\sum_{i,j} \left(
		\Delta^{-1}_{I\,ii}\, \Delta_{I\,ij}\,  \Delta^{-1}_{I\,jj}
		-\Delta^{-1}_{I\,ij} \Delta_{I\,ij}\,  \Delta^{-1}_{I\,ij}
	\right)\, .
\end{equation*}
Recall that $\Delta_I$ is the Laplacian matrix restricted to inner vertices.

In the result (\ref{eq:covariant_eff_action}) for the free energy, we find, next to the Yang-Mills action for the gauge field, also the intrinsic curvature scalar $\bar R$ on $\cS$ and further curvature terms and quadratic terms in the second fundamental form $B$. 
All coefficients are expectation values of graph invariants, where the coefficients $a_1,\ldots, a_3$ are not determined explicitly. Rather, using the normal coordinates according to Graham and Kuo \cite{Graham:2024gnc}, one can convince oneself that, to first order in $\alpha$, these are the only allowed curvature terms, and the coefficients must be given by expectation values of graph invariants. 

\section{Concluding remarks}
\label{sec:concluding_remarks}

We have analyzed a non-perturbative definition of (discrete) string theory in terms of a matrix model on a higher-dimensional Riemannian manifold with metric and gauge field background. The action of the matrix model has a very simple dependence on the background fields, but still leads directly to the Einstein-Hilbert and Yang-Mills actions. The cosmological and gravitational constants are determined by the graph combinatorics of ribbon graphs, encoded in terms of invariant graph polynomials in the formal power series $\hat w$.
Quite remarkably, the resulting expressions for the cosmological and gravitational constants contain all orders in string perturbation expansion.

The derivation of the free energy did not require the on-shell condition for the metric and the gauge field. This is different from the continuum formulation of perturbative string theory, where the effective action is deduced indirectly from the equations of motion. The latter are a consequence of quantization and requiring Weyl invariance.
In the discrete setting there is no Weyl invariance, but we found that the coefficient of the curvature scalar term in the free energy depends on the expectation values of the vertex and edge numbers, thus changing with graph size. This is in line with the expectation that going off-shell comes with scale dependent terms.

In this work the metric was considered as a background field, but in principle we could consider integrating over the metric.
Our result (\ref{eq:free_energy_curved}) could then be interpreted as an effective action for the metric. Integrating out the matrix degrees of freedom is justified for small values of $\alpha$, when the matrix interactions are localized to small distances as compared to typical changes of the metric.

Since the metric is included as a background field in the matrix model, gravity is not an emerging phenomenon. This is different from some examples of large $N$ duality, where gravity is indeed emerging dynamically: for instance, the polarized IKKT matrix model studied in \cite{Komatsu:2024bop,Komatsu:2024ydh} or generally the AdS/CFT correspondence \cite{Maldacena:1997re}, where supergravity on AdS${}_{D+1}$ is holographically emerging from the dual CFT${_D}$. Notice that in the AdS/CFT context, a technically very similar approach to the present one, based on heat kernel propagators, was taken in \cite{Gopakumar:2003ns} to relate tree amplitudes on both sides of the duality.

Let us close with some topics that were not considered in this article, but would be worth looking at, based on our results and in addition to them. 

There were no fermionic fields considered, indeed we restricted our attention to bosonic backgrounds fields only. 
Furthermore, our focus was on the partition function and the free energy. It might be interesting to study vertex operators and derive results for their amplitudes. 

Although we proposed a nonperturbative definition of a (discrete) string theory through a matrix model, our eventual considerations were restricted to perturbative results. The methods of resurgence and transseries of \cite{Marino:2008ya,Pasquetti:2010bps,Marino:2012zq,Aniceto:2018bis} might unlock the full non-perturbative potential.

All analysis was done for Riemannian signature metrics, not considering time-like coordinates. Let us have some speculative remarks on how to introduce a time coordinate.

Wick rotation is a common trick in the path integral approach to quantum field theories for switching between time-like and space-like coordinates. But since the kinetic term of the matrix model is exponential, that is $e^{-\alpha/2\, \Delta}$, a naive Wick rotation of one of the coordinates would lead to a non-integrable Green's function due to a factor 
$e^{t^2/2\alpha}$.

Indeed, the exponential behavior of the Green's function can be made well-defined for a Minkowski signature metric, if the exponent is imaginary rather than real.
To get there, we analytically continue both, the time coordinate and the space coordinates and perform a $1/2$-Wick rotation, so that the Green's function formally turns into a solution of a ``Schrödinger'' equation instead of the heat equation.
More explicit, let us rotate the coordinate $x^0$ into a time coordinate and $x^a$ for $a=1,\ldots,D$ into space coordinates, by introducing a rotation angle 
$\varphi \in [0,\pi/4]$ and 
$$
  x^0 \rightarrow e^{i\varphi} t\, ,\quad \textrm{and}\quad
	x^a \rightarrow e^{-i\varphi} x^a\, . 
$$
Considering the Gaussian integration within the bounds for $\varphi$, we get for the time coordinate
$$
  \int_{-\infty}^{\infty} dt e^{i\varphi} 
	e^{- e^{2i\varphi}\frac{t^2}{2\alpha}} = \sqrt{2\pi\alpha}, 
$$
and for the space coordinates
$$
  \int_{-\infty}^{\infty} dx e^{-i\varphi}
	e^{- e^{-2i\varphi}\frac{x^2}{2\alpha}} = \sqrt{2\pi\alpha}. 
$$

The full rotation to $\varphi = \pi/4$ then leads to the Green's function
$$
  G(x,x') =
	\frac{1}{(2\pi \alpha)^{D/2}} e^{i\frac{-(t-t')^2 + (x-x')^2}{2\alpha}}\, .
$$
In such a way, the results for the matrix model with Laplacian operator for the Euclidean signature metric can be continued to results for a Minkowski signature metric by applying the $1/2$-Wick rotation
$$
  x^0 \rightarrow \sqrt{i}\, t\, ,\quad \textrm{and}\quad
	x^a \rightarrow \sqrt{-i}\, x^a\, . 
$$
Note however that this rotation gives rise to a prefactor $i^{1-D/2}$ in the integration measure of the action, whose interpretation is unclear in the free energy.

There is yet another approach to obtaining a time-like dimension from the matrix model. From the study of matrix models for $D \leq 1$ it is known from different angles \cite{Das:1990kaa, Kostov:1999xi,Ginsparg:1993is} that the matrix degrees of freedom form an additional spacetime dimension in the double scaling limit, described by a Liouville field.

However, as discussed in Section \ref{sec:MMAD}, the boundary condition for the matrix field requires a vanishing field at infinity and thus does not allow constant modes. These constant modes are (by definition) present in the $D < 1$ matrix models. For $D=1$ matrix models, considered in \cite{Das:1990kaa,Ginsparg:1993is}, it was argued that both kinetic terms, $e^{-\partial_x^2}$ and $\partial_x^2$, lead to the same critical behavior in the double scaling limit. The kinetic term $\partial_x^2$ does allow constant modes though.

Because of the lack of constant modes for the exponential kinetic term, it is unclear to the author whether the matrix degrees of freedom do generate an additional Liouville dimension, or whether it is necessary to explicitly couple a constant matrix to the matrix field $M(x)$ to achieve this goal.
Additionally, it is not clear how general covariance can be shown or made explicit for an emergent time coordinate in the matrix model with curved background.

\newpage


\appendix

\section{The free partition function}
\label{apdx:free_partition_function}

The free partition function for the matrix model is
$$
  Z_\ni(\alpha,\lambda,N) = \int{\cD M(x)\; 
	\exp{\left(-\frac{N}{2 \lambda} \int \frac{d^Dx}{(2\pi\alpha)^{D/2}}
	\trace M(x) e^{-\frac{\alpha}{2} \Delta} M(x) \right)}}\, .
$$
One way to define this functional integral is to discretize space on a hypercube in the lattice $\mathbb{Z}^D$ with spacing $\Delta x$ and to require an appropriate normalization condition for the measure. 
For a scalar field $m(x)$ the integration measure shall be normalized so that
\begin{equation*}
 \int{\prod_x\cD m(x)\; 
	exp{\left(-\frac{1}{2} \int \frac{d^Dx}{(2\pi\alpha)^{D/2}}
	m(x)^2 \right)}} = 1\, .
\end{equation*}

Let us take the edge length of the hypercube in $\bbZ^D$ to be $2K$ for a large integer $K$ and discretize as follows:
$$
  \int d^Dx f(x) \sim 
	\Delta x^D \sum_{\mathbf{a}} f_\bfa := 
	\Delta x^D \sum_{a_1\ldots a_D=-K+1}^K f_\bfa
	\, .
$$
The volume is given by $V_\cM = (2K)^D \Delta x^D$. 

The Dirac delta distribution becomes
$$
  \delta^D(x-x') \sim \frac{1}{\Delta x^D} \delta_{\bfa,\bfa'}\, .
$$
In particular, $\delta^D(0) \sim \frac{1}{\Delta x^D}$.

The discrete functional integral measure is defined by:
$$
  \prod_x\cD m(x) \sim 
	\frac{1}{C} \prod_{\mathbf{a}} d m_\mathbf{a} 
	\, .
$$

With these ingredients the normalization condition for the functional integral is
$$
  \frac{1}{C} 
	\int\prod_{\mathbf{a}} dm_\mathbf{a}
	\exp{\left(
		-\frac{1}{2} 
		\frac{\Delta x^D}{(2\pi\alpha)^{D/2}}
		\sum_{\mathbf{a}} m_\mathbf{a}^2
	\right)} = 1\, , 
$$
and the normalization constant $C$ is determined to be
$$
  C = \left(
	      2\pi \frac{(2\pi\alpha)^{D/2}}{\Delta x^D}
			\right)^{\frac{(2K)^D}{2}} 
			\sim \left(
			  2\pi (2\pi\alpha)^{D/2}\delta^D(0)
			\right)^{\frac{V_\cM\delta^D(0)}{2}}
$$

To compute the functional integral of the free partition function, the Laplacian operator must be discretized as well:
$$
  \Delta_x f(x) \sim 
	\frac{1}{\Delta x^2}\hat\Delta_{\bfa\bfb} f_\bfb := 
	\frac{1}{\Delta x^2}
	\sum_{d=1}^D f_{\bfa+\bfe_d} - 2 f_\bfa + f_{\bfa-\bfe_d}\, ,
$$ 
where $\bfe_d$ is a $D$-tuple with $1$ at the $d$th position and zeros otherwise.
$\hat\Delta_{\bfa\bfb}$ is a $(2K)^D$-dimensional matrix with $-2D$ at the diagonal.

Now, all preparations are done to compute the discretized functional integral
\begin{eqnarray*}
  Z_\ni(\alpha,\lambda,N) &\sim& 
	\frac{1}{C^{N^2}} \int\prod_{ij,\mathbf{a}} dM_{ij\mathbf{a}}
	\exp{\left(
		-\frac{N}{2\lambda} \frac{\Delta x^D}{(2\pi\alpha)^{D/2}}
		\sum_{\mathbf{ab}} 
		M_{ij,\mathbf{a}} \bigl(e^{-\frac{\alpha}{2\Delta x^2} \hat\Delta}\bigr)_{\bfa\bfb} 
		M^*_{ij,\mathbf{b}}
	\right)} = \\
	&=&
	\left(
	  \int\prod_{\mathbf{a}}
		\frac{dm_{\mathbf{a}}}{\sqrt{2\pi}}
		e^{
			-\frac{N}{2\lambda} \sum_{\mathbf{ab}}
			m_{\mathbf{a}} \bigl(e^{-\frac{\alpha}{2\Delta x^2} \hat\Delta}\bigr)_{\bfa\bfb} m_{\mathbf{b}}
		}
	\right)^{N^2} = \\
	&=&
	\left(
	  (\lambda/N)^{(2K)^D} 
	  \det (e^{\frac{\alpha}{2\Delta x^2}\hat\Delta} )
	\right)^{\frac{N^2}{2}} = \\
	&=&  
	\left(\lambda/N \right)^{\frac{N^2}{2}(2K)^D}
	~ e^{-\frac{N^2}{2} \frac{\alpha}{\Delta x^2} D(2K)^D}
\end{eqnarray*}

Using that the volume of the discretized space is $(2K)^D = V_\cM \Delta x^{-D} \sim V_\cM \delta^D(0)$, the Gaussian part of the partition function becomes
\begin{equation}
	\label{eq:NonIntPartition}
	Z_\ni(\alpha,\lambda,N) = 
	\left(\lambda/N \right)^{\frac{N^2}{2}V_\cM\delta^D(0)} ~ 
	e^{-\frac{N^2}{2} \alpha D V_\cM \delta^{D+2}(0)}\, .
\end{equation}

Expressed in terms of the free energy this becomes
$$
  \cF_\ni = - \left(
	  \frac{N^2}{2} \ln\left(\lambda/N \right) 
	  - \alpha D \delta^2(0)
	\right)
	V_\cM \delta^D(0)\, .
$$

\section{The heat kernel in curved background}
\label{apdx:heat_kernel_curved}

In this section we integrate out the discrete embedding coordinates $X$ in (\ref{eq:free_energy_curved_gaussian}) when the matrix model is considered on a Riemannian manifold $\cM$. The restrictions on $\cM$ as introduced in Section \ref{sec:MMCB} shall apply.

Let us consider the skeleton $\gamma$ of a ribbon graph $\Gamma$ without boundary components, which is uniquely (up to similarity transformations) determined by the adjacency matrix $A = A(\gamma)$. The contribution to the perturbative free energy from the heat kernels (\ref{eq:curved_Greens_function}) is given in leading order in $\alpha$ by
\begin{equation}
  \label{eq:pert_heat_kernels}
  \int_\cM \prod_{i=1}^v \frac{d^Dx_i \sqrt{g(x_i)}}{(2\pi\alpha)^{D/2}}  
  \prod_{e_{ij}} \sqrt{\bfD(x_i, x_j)} 
	e^{-\frac{1}{\alpha} \sigma(x_i,x_j)} 
	\left(1 + \frac{\alpha}{12} R(x_j) \right) \,
\end{equation}
where $\bfD$ is the Van Vleck-Morette determinant (\ref{eq:VVM_determinant}).

Just as in Euclidean space we expand this expression around a reference point, which we choose as one of the embedding coordinates, say $x = x_v$ \cite{Carneiro:2024slt}, and introduce $x'_i = x_i - x$ for $i=1,\ldots,v-1$. In Riemannian normal coordinates around $x$ (not using the full freedom to write the metric at $x$ as Kronecker delta, but keeping $g_{\mu\nu} = g_{\mu\nu}(x)$) the metric can be expanded as
\begin{eqnarray*}
  g_{\mu\nu}(x_i) &=& g_{\mu\nu}(x) - 
	\frac{1}{3} R_{\mu\rho\nu\sigma}(x)x'_i{}^\rho x'_i{}^\sigma + \cO(x'_i{}^3)\,  , \\
	\sqrt{\det{g_{\mu\nu}(x_i)}} &=& \sqrt{\det{g_{\mu\nu}(x)}}
	\left(
		1 - \frac{1}{6} R_{\rho\sigma}(x) x'_i{}^\rho x'_i{}^\sigma
	\right) + \cO(x'_i{}^3)\, .
\end{eqnarray*}
Half the geodesic square distance $\sigma(x_i,x_j)$ and the van Vleck-Morette determinant can be expanded as \cite{Barvinsky:2021ijq}
\begin{eqnarray}
	\label{eq:Morette_expansion}
	\nonumber
	\sigma(x_i,x_j) &=& 
	\frac{1}{2} g_{\mu\nu}(x) (x'_i-x'_j)^\mu(x'_i-x'_j)^\nu -
	\frac{1}{6} R_{\mu\rho\nu\sigma}(x) x'_i{}^\mu x'_i{}^\nu x'_j{}^\rho x'_j{}^\sigma + \cO(x'_i{}^5)\, , \\
	\nonumber
	\sigma(x_i,x) &=& 
	\frac{1}{2} g_{\mu\nu}(x) x'_i{}^\mu x'_i{}^\nu\, \\
	\sqrt{\bfD(x_i, x_j)} &=& 1 + \frac{1}{12} R_{\mu\nu}(x) (x_i-x_j)^\mu(x_i-x_j)^\nu + \cO(x'_i{}^3)\, , \\
	\nonumber
	\sqrt{\bfD(x_i, x)} &=& 1 + \frac{1}{12} R_{\mu\nu}(x) x_i{}^\mu x_i{}^\nu + \cO(x'_i{}^3)\, .
	\nonumber
\end{eqnarray}
Only terms that eventually lead to first order terms in $\alpha$ are kept, and (\ref{eq:pert_heat_kernels}) becomes
\begin{eqnarray}
  \label{eq:curved_heat_kernel_expanded}
  && \int_\cM \frac{d^Dx \sqrt{g}}{(2\pi\alpha)^{D/2}} \int \frac{d^DX' \sqrt{g}^{v-1}}{(2\pi\alpha)^{D(v-1)/2}}\,
  \exp{\left(-\frac{1}{2\alpha} g_{\mu\nu} X'^\mu \Delta' X'^\nu\right)} \times \\
	&\times& \left(
	  1 + \frac{1}{12} 
		\left(
			\alpha e R + R_{\mu\nu} X'^\mu \Delta' X'^\nu -
			2 R_{\mu\nu} X'^\mu X'^\nu
			- \sum_{i,j} \frac{1}{\alpha} R_{\mu\rho\nu\sigma} 
			x'_i{}^\mu x'_i{}^\nu \Delta'_{ij} x'_j{}^\rho x'_j{}^\sigma
		\right)
	\right) \nonumber
\end{eqnarray}
where $\Delta'$ is the reduced Laplacian matrix for $\gamma$ and $X' = (x'_1, \ldots, x'_{v-1})$. The metric and its curvature are understood to be evaluated at $x$. The integration over $\int d^DX'$ gives 
\begin{equation*}
  \frac{1}{(\det{\Delta'})^{D/2}} 
  \int_\cM \frac{d^Dx \sqrt{g}}{(2\pi\alpha)^{D/2}} 
	\left(
		1 + \frac{\alpha}{12}
		\left(
			e + v - 1 - \cI
		\right) R(x) + \cO(\alpha^2)
	\right)\, .
\end{equation*}
The final result of the above computation must be independent of the choice of the reference point $x$, which means in particular that $\cI = \cI(\gamma)$ must be a graph invariant. It originates from the last two terms in (\ref{eq:curved_heat_kernel_expanded}) and is given by
\begin{equation*}
  \cI(\gamma) = 2 \sum_i \Delta'^{-1}_{ii} + 
	\sum_{i,j} \left(
		\Delta'^{-1}_{ii}\, \Delta'_{ij}\,  \Delta'^{-1}_{jj}
		-\Delta'^{-1}_{ij} \Delta'_{ij}\,  \Delta'^{-1}_{ij}
	\right)\, .
\end{equation*}

\section{The heat kernel along a Riemannian submanifold}
\label{apdx:heat_kernel_submfld}

Let $\cS$ be a $d$-dimensional Riemannian submanifold of a $D$-dimensional Riemannian manifold $\cM$. For a given skeleton $\gamma$ with inner and boundary vertices as introduced in Section \ref{sec:VMM} the leading contribution in $\alpha$ to the free energy of the matrix model (\ref{eq:VMM}) has the following heat kernels:
\begin{eqnarray*}
  \prod_i \int_\cM d^Dx_i \sqrt{g(x_i)}\hspace{-20pt}
	&& \prod_p \int_\cS d^dy_i \sqrt{\bar g(y_p)} ~ 
	\prod_{e_{ij}} \sqrt{\bfD(x_i, x_j)} 
	e^{-\frac{1}{\alpha} \sigma(x_i,x_j)} 
	\left(1 + \frac{\alpha}{12} R(x_j) \right) \\
	&\times&\prod_{e_{ip}} \sqrt{\bfD(x_i, y_p)} 
	e^{-\frac{1}{\alpha} \sigma(x_i,y_p)} 
	\left(1 + \frac{\alpha}{12} R(y_p) \right) \\
	&\times&\prod_{e_{pq}} \sqrt{\bar \bfD(y_p,y_q)} 
	e^{-\frac{1}{\alpha} \sigma(y_p,y_q)}  W(y_p,y_q)
	\left(1 + \frac{\alpha}{12} \bar R(y_q) + 
	  \frac{\alpha^2}{48} F _{\alpha\beta} F^{\alpha\beta}(y_q)
	\right)\, .
\end{eqnarray*}
Here $\bar g = \iota^*g$ is the induced metric on $\cS$. $\bar \bfD(y_p, y_q)$ and $\bar R(y)$ are the Van Vleck-Morette determinant and the Ricci scalar for the induced metric, respectively.

As argued in the main text the integrants localize in a tubular neighborhood of $\cS$ in $\cM$, the size controlled by $\alpha$.
We use normal coordinates as introduced by Graham and Kuo \cite{Graham:2024gnc}, expanding around a reference point, e.g.~$y_{v_B} = y$ on $\cS$, so that the expansion coefficients are given by polynomials of the curvature tensor $R$ and the second fundamental form $B$. 
Recall that the normal coordinates are chosen as $x^\mu = (y^\alpha,z^{\alpha'})$ with $y$ being the coordinates along $\cS$ and $z$ along $\cN\cS$.
The metric components in tangent and normal direction are given to first order in $R$ and second order in $B$ by
\begin{eqnarray*}
  g_{\alpha\beta}(x_i) &=& \bar g_{\alpha\beta}(y) 
	- 2 B_{\alpha\beta\alpha'}(y)~ z_i^{\alpha'}
	-\frac{1}{3} \bar R_{\alpha\beta\gamma\delta}(y)~ y'_i{}^\gamma y'_i{}^\delta
	- R_{\alpha\alpha'\beta\beta'}(y)~ z_i^{\alpha'}z_i^{\beta'}\, ,\\
	g_{\alpha\beta'}(x_i) &=& 
	\frac{1}{2} \bigl(R_{\alpha\beta\alpha'\beta'}(y) 
	- B_{\beta'\alpha}{}^\gamma B_{\alpha'\beta \gamma}
	+ B_{\beta'\beta}{}^\gamma B_{\alpha'\alpha \gamma} \bigr)~ 
	y'_i{}^\beta z_i^{\alpha'}
	+ \frac{2}{3} R_{\alpha\gamma'\delta'\beta'}(y)~ z_i^{\gamma'} z_i^{\delta'}\, , \\
	g_{\alpha'\beta'}(x_i) &=& g^\perp_{\alpha'\beta'}(y)
	- \frac{1}{3} R_{\alpha'\beta'\gamma'\delta'}(y)~ z_i^{\gamma'} z_i^{\delta'} \, ,
\end{eqnarray*}
where $y'_i = y_i - y$. We use these coordinates and the Gauss equation (\ref{eq:Gauss_equation}) to expand the heat kernels and metric determinants, explicitly keeping the curvature $\bar R_{\alpha\beta\gamma\delta}$ of the induced metric, the curvature normal component $R_{\alpha'\beta'\gamma'\delta'}$ and the gauge field, recall the definitions (\ref{eq:extrinsic_curv_defs}). (Mixed curvature components, although present, are not in focus in this work.)

The small distance expansion of the Van Vleck-Morette determinant \cite{Visser:1992pz} gives
\begin{eqnarray*}
  \sqrt{\bfD(x_i,x_j)} &=& 1 + 
	\frac{1}{12} \bar R_{\alpha\beta}(y) y'^\alpha_{ij}y'^\beta_{ij} + 
	\frac{1}{12} \bar R^{\perp\perp}_{\alpha'\beta'}(y) z^{\alpha'}_{ij}z^{\beta'}_{ij} + 
	\ldots\, ,\\
	\sqrt{\bfD(x_i,y_q)} &=& 1 + 
	\frac{1}{12} \bar R_{\alpha\beta}(y) y'^\alpha_{iq}y'^\beta_{iq} + 
	\frac{1}{12} \bar R^{\perp\perp}_{\alpha'\beta'}(y) z^{\alpha'}_{i}z^{\beta'}_{i} + \ldots\, ,\\
	\sqrt{\bfD(x_i,y)} &=& 1 + 
	\frac{1}{12} \bar R_{\alpha\beta}(y) y'^\alpha_{i}y'^\beta_{i} + 	
	\frac{1}{12} \bar R^{\perp\perp}_{\alpha'\beta'}(y) z^{\alpha'}_{i}z^{\beta'}_{i} + \ldots\, ,\\
	\sqrt{\bar\bfD(y_p,y_q)} &=& 1 + 
	\frac{1}{12} \bar R_{\alpha\beta}(y) y'^\alpha_{pq}y'^\beta_{pq} + \cO(\partial \bar R)\, ,\\
	\sqrt{\bar\bfD(y_p,y)} &=& 1 + 
	\frac{1}{12} \bar R_{\alpha\beta}(y) y'^\alpha_{p}y'^\beta_{p} + \cO(\partial \bar R)\, ,
\end{eqnarray*}
where $y'_{ij} = y'_i - y'_j$, and the dots $\ldots$ are short for $\cO(R^{\parallel\perp}y^2, B^2y^2,Ryz)$ and higher derivative orders of the curvature.
The metric determinant is expanded as
\begin{eqnarray*}
  \sqrt{g(x_i)} &=& 
	\sqrt{\bar g(y)} \sqrt{g^\perp(y)} 
	\left(1 - \frac{1}{6} \bar R_{\alpha\beta}(y) y'^\alpha_{i}y'^\beta_{i} 
	- \frac{1}{6} \bar R^{\perp\perp}_{\alpha'\beta'}(y) z^{\alpha'}_{i}z^{\beta'}_{i} 
	+ \ldots \right)\, ,
	\\
	\sqrt{\bar g(y_p)} &=& 	
	\sqrt{\bar g(y)} 
	\left(1 - \frac{1}{6} \bar R_{\alpha\beta}(y) y'^\alpha_{p}y'^\beta_{p} + \cO(\partial \bar R) \right)\, .
\end{eqnarray*}
And the expansion of half the geodesic distance square $\sigma$ is
\begin{eqnarray*}
  \sigma(x_i,x_j) &=& 
	\frac{1}{2} \bar g_{\alpha\beta}(y) y'{}_{ij}^{\alpha} y'{}_{ij}^{\beta} -
	\frac{1}{6} \bar R_{\alpha\gamma\beta\delta}(y)
	y'{}_i^\alpha y'{}_i^\beta y'{}_j^\gamma y'{}_j^\delta + \\ 
	&+& \frac{1}{2} g^\perp_{\alpha'\beta'}(y) z_{ij}^{\alpha'} z_{ij}^{\beta'} - 
	\frac{1}{6} R^{\perp\perp}_{\alpha'\gamma'\beta'\delta'}(y) 
	z_i^{\alpha'} z_i^{\beta'} z_j^{\gamma'} z_j^{\delta'} +\ldots,\\
  \sigma(x_i,y_q) &=& 
	\frac{1}{2} \bar g_{\alpha\beta}(y) y'{}_{iq}^{\alpha} y'{}_{iq}^{\beta} -
	\frac{1}{6} \bar R_{\alpha\gamma\beta\delta}(y) 
	y'{}_i^\alpha y'{}_i^\beta y'{}_q^\gamma y'{}_q^\delta +
	\frac{1}{2} g^\perp_{\alpha'\beta'}(y) z_{i}^{\alpha'} z_{i}^{\beta'} + \ldots,\\
  \sigma(x_i,y) &=& 
	\frac{1}{2} \bar g_{\alpha\beta}(y) y'{}_{i}^{\alpha} y'{}_{i}^{\beta} + 
	\frac{1}{2} g^\perp_{\alpha'\beta'}(y) z_{i}^{\alpha'} z_{i}^{\beta'} + \ldots,\\
  \bar\sigma(y_p,y_q) &=& 
	\frac{1}{2} \bar g_{\alpha\beta}(y) y'{}_{pq}^{\alpha} y'{}_{pq}^{\beta} -
	\frac{1}{6} \bar R_{\alpha\gamma\beta\delta}(y) 
	y'{}_p^\alpha y'{}_p^\beta y'{}_q^\gamma y'{}_q^\delta + \ldots,\\
  \bar\sigma(y_p,y) &=& 
	\frac{1}{2} \bar g_{\alpha\beta}(y) y'{}_{p}^{\alpha} y'{}_{p}^{\beta} + \ldots
	\, .
\end{eqnarray*}
Collecting all contributions the integral over the heat kernels is then approximated by
\begin{eqnarray}
  \label{eq:int_over_heat_kernels}
  \int_\cS d^dy \sqrt{\bar g} && \hspace{-25pt}
	\int_{T\cS|_y} d^{d} Y' 
	\sqrt{\bar g}^{v-1} 
	e^{-\frac{\bar g_{\alpha\beta} Y'^\alpha \Delta' Y'^\beta}{2\alpha}}
	\int_{\cN\cS|_y} d^{D-d} Z \sqrt{\bar g^\perp}^{v_I}
	e^{-\frac{g^\perp_{\alpha'\beta'} Z^{\alpha'} \Delta_I Z^{\beta'}}{2\alpha}}
	 \times
	\\
	\times \biggl(
		1 \hspace{-5pt} \biggr. &+&\hspace{-5pt} \frac{\alpha}{12}\Bigl(
			\alpha e \bar R + \bar R_{\alpha\beta} Y'^\alpha \Delta' Y'^\beta -
			2 \bar R_{\alpha\beta} Y'^\alpha Y'^\beta
			- \sum_{i,j} \frac{1}{\alpha} \bar R_{\alpha\gamma\beta\delta} 
			y'_i{}^\alpha y'_i{}^\beta \Delta'_{ij} y'_j{}^\gamma y'_j{}^\delta
		\Bigr)  \nonumber \\
		&+&\hspace{-5pt} \frac{\alpha}{12}\Bigl(
			\alpha e_I R^{\perp\perp} 
			\hspace{-3pt} + \hspace{-3pt}
			R^{\perp\perp}_{\alpha'\beta'} Z^{\alpha'} \Delta_I Z^{\beta'} 
			\hspace{-3pt} - \hspace{-3pt}
			2 R^{\perp\perp}_{\alpha'\beta'} Z^{\alpha'} Z^{\beta'}
			\hspace{-3pt} - \hspace{-3pt}
			\sum_{i,j} \frac{1}{\alpha} R^{\perp\perp}_{\alpha'\gamma'\beta'\delta'} 
			z_i^{\alpha'} z_i^{\beta'} \Delta_I{}_{ij} z_j^{\gamma'} z_j^{\delta'}
		\Bigr)  \nonumber \\
		&+&\hspace{-5pt} \biggl.\frac{\alpha^2 e_B}{48 n} \trace{F^2(y)} 
		+ \ldots \biggr)
		\, . \nonumber
\end{eqnarray}
Here, $e_I$ is the number of inner edges, and $\Delta_I$ is the Laplacian matrix restricted to inner vertices.

\end{document}